\documentclass[fleqn,usenatbib]{mnras}

\usepackage{mathtools}

\usepackage[T1]{fontenc}

\DeclareRobustCommand{\VAN}[3]{#2}
\let\VANthebibliography\thebibliography
\def\thebibliography{\DeclareRobustCommand{\VAN}[3]{##3}\VANthebibliography}

\usepackage{graphicx}	
\usepackage{amsmath,amsfonts,amssymb,epsfig}
\usepackage{algorithm2e}
\usepackage{booktabs}
\usepackage{subfig, subfloat}

\title[Point Process Model of M33 Star Clusters]
{Gibbs Point Process Model for Young Star Clusters in M33}

\author[Li \& Barmby]{
Dayi Li,$^{1,2}$\thanks{Email: dayi.li@mail.utoronto.ca}
Pauline Barmby$^{2,3}$
\\
$^{1}$Department of Statistical Sciences, University of Toronto,
100 St. George St, Toronto, ON M5S 3G3, Canada \\
$^{2}$Department of Statistical and Actuarial Sciences \& School of Mathematical and Statistical Sciences,\\
Western University,
1151 Richmond St,
London, ON N6A 3K7, Canada\\
$^{3}$Department of Physics and Astronomy \&  Institute for Earth \& Space Exploration, Western University
}

\date{Accepted XXX. Received YYY; in original form ZZZ}

\pubyear{2020}

\begin{document}
\label{firstpage}
\pagerange{\pageref{firstpage}--\pageref{lastpage}}
\maketitle

\begin{abstract}
We demonstrate the power of Gibbs point process models from the spatial statistics literature when applied to studies of resolved galaxies. We conduct a rigorous analysis of the spatial distributions of objects in the star formation complexes of M33, including giant molecular clouds (GMCs) and young stellar cluster candidates (YSCCs). We choose a hierarchical model structure from GMCs to YSCCs based on the natural formation hierarchy between them. This approach circumvents the limitations of the empirical two-point correlation function analysis by naturally accounting for the inhomogeneity present in the distribution of YSCCs. We also investigate the effects of GMCs' properties on their spatial distributions. We confirm that the distribution of GMCs and YSCCs are highly correlated. We found that the spatial distributions of YSCCs reaches a peak of clustering pattern at $\sim 250$~pc scale compared to a Poisson process. This clustering mainly occurs in regions where the galactocentric distance $\gtrsim 4.5$~kpc. Furthermore, the galactocentric distance of GMCs and their mass have  strong positive effects on the correlation strength between GMCs and YSCCs.
We outline some possible implications of these findings for our understanding of the cluster formation process.
\end{abstract}

\begin{keywords}
star formation, galaxies: individual: M33, galaxies: ISM, galaxies: star clusters, Methods: statistical
\end{keywords}




\section{Introduction}

\label{sec:intro}

The spatial distributions of star clusters (SCs) and giant molecular clouds (GMCs), as well as their spatial relationships, provide crucial information in understanding the star formation process. Star formation is generally considered to take place within GMCs \citep{kennicutt_star_2011}. The distribution of star formation is understood to be resulting from GMC fragmentation \citep{carlberg_magnetic_1990, mclaughlin_formation_1996}, under the influence of gas collapse under immense gravitational effects \citep{vega_self-gravity_1996, kuznetsova_kinematics_2018}, turbulence in local environment \citep{elmegreen_interstellar_2004, federrath_fractal_2009, girichidis_importance_2012, hopkins_meaning_2013, guszejnov_protostellar_2017} or feedback processes that suppress the star formation \citep{krumholz_big_2014}. 

Investigating the spatial distribution of SCs provides a sensitive and direct observational signature of the star formation process \citep{grasha_spatial_2019}. However, it is not well understood to what extent the galactic environment, locally and globally, influences the evolution of SCs \citep{grasha_spatial_2019}. Understanding and measuring the spatial distribution  is then a crucial task. One current method used in understanding this distribution is called the two-point correlation function (2PCF) in astronomical literature \citep{peebles_large-scale_1980} or the pair correlation function (PCF) in spatial statistics literature \citep{baddeley_spatial_2015}. It measures the excess probability of finding two objects at a certain distance away compared to that for a completely random Poisson distribution of objects. 
The 2PCF was first derived by \cite{peebles_large-scale_1980}  for use in studying the large scale structure of the Universe. In the spatial statistics literature \citep{cressie_statistics_2011, baddeley_spatial_2015}, it took up the name pair correlation function (PCF) due to its origin in statistical mechanics for studying the distributional structure of molecules in complex $N$-body systems. The two functions are exactly the same except for different normalizing conventions. In this paper, the terms will be used interchangeably depending on the context. 

As noted in both the spatial statistics and astronomy literature \citep{moller_statistical_2003, baddeley_spatial_2015,peebles_large-scale_1980, peebles_galaxy_2001}, a crucial assumption on the validity of the 2PCF is that the point pattern has to be homogeneous and stationary. Homogeneous here means the projected number density  of the point pattern is constant in all regions and stationary means the point pattern is translation invariant, i.e., the distribution of the point pattern does not change as the position of the observational reference point is shifted.  Stationarity implies homogeneity: in fact homogeneity can be regarded as first-order stationarity, i.e., the chance of individual points occurring does not change as one shifts location.  Second-order stationary means that the relationship between pair of points does not depend on the absolute positions of the points but only on their relative positions. This assumption is generally assumed for most point patterns analysed.
For analysing the large scale structure of the Universe, there has been accumulating evidence supporting the claims of homogeneity and stationarity \citep{peebles_principles_1993, davis_galaxy-weighted_1997, peebles_galaxy_2001} of the galaxy distribution on the scales of $10 \sim 200 h^{-1}$Mpc. Therefore, the application of 2PCF in this context is justified.

When the assumption of homogeneity is violated, i.e., when there is evidence of inhomogeneity, empirical 2PCF/PCF is no longer a valid tool for interpoint interaction analysis. It is important to highlight the subtle difference between the effect from inhomogeneity and the interpoint interaction. Inhomogeneity, usually arising from external effects, exerts its influence on the occurrence of a point independently of another point. We can think of this as a ``fertility" effect \citep{baddeley_spatial_2015}, i.e., how much resource there is in a certain region to produce one point. The interpoint interaction, however, is the influence exerted from the occurrence of a point to another point, i.e., there exists a notion of dependence structure. We can think of this as competition of resources in the case of inhibition or triggering of occurrences of multiple points in the case of clustering \citep{baddeley_spatial_2015}. Therefore, excluding the effects of inhomogeneity can lead to drastic differences in conclusions from a fitted 2PCF. 

In the context of stellar population studies, the aim of 2PCF is to measure the interpoint interaction effect. This means that the violation of homogeneity can lead to drastically different conclusions from the fitted 2PCF. For example, projected mass and star cluster number density distributions in galaxies decline with galactocentric distance.  Fitting an empirical 2PCF/PCF to the star cluster distribution  without considering this confounding factor is unlikely to provide an unbiased estimate of the actual distribution pattern of SCs since they are already exhibiting a clustering pattern. To account for inhomogeneity from external effects is problematic if 2PCF is the only tool we have. For our data, there is no numerical measurement of inhomogeneous external effects so that we can eliminate them and refit the 2PCF.

Recently, the empirical 2PCF and its variant have been applied to analyse the spatial distribution of SCs by \citet{grasha_spatial_2015, grasha_hierarchical_2017, corbelli_molecules_2017, grasha_spatial_2019}. However, due to the limitations of 2PCF in dealing with highly inhomogeneous point pattern such as SCs, the questions that these previous studies are able to answer are limited and the obtained conclusions can be potentially biased.
In their studies of star cluster spatial distributions,
\citet{corbelli_molecules_2017} and \citet{ grasha_spatial_2019} attempted to address the issue of inhomogeneity due to the large scale variation in  number density across the galaxy disc, but used a method that is rather ad-hoc and prone to information loss. They chose to separate the galactic plane into several annuli encompassing the galaxy centre so that the large scale variation could be regarded as homogeneous in each region. However, there is no guarantee that the distribution of points in each region is homogeneous since there might also exist local inhomogeneity. Grouping the data also introduces information loss since information on a continuous space is cut into several non-communicating subspaces.
Another limitation of 2PCF is its restriction on investigating how the properties of SCs and GMCs affect their spatial relationships. \cite{grasha_spatial_2015, grasha_spatial_2019, grasha_hierarchical_2017} investigated the effect of age and mass on the clustering strength of SCs. The data has to be grouped by age and mass to provide an analysis from the 2PCF. This grouping of data loses significant amount of information since a continuous variable is reduced to a categorical variable.

Building on these previous studies, we introduce a parametric modelling approach --- Gibbs point process \citep{ripley_markov_1977, baddeley_spatial_2015} --- to circumvent the limitations of 2PCF and analyse the spatial structure of SCs in M33. GPP models are ubiquitous for modelling point patterns where interpoint interaction is considered. Originating from statistical physics, these models were first employed to study the behaviour of physical systems with massive numbers of interacting particles exhibiting a complex dependence structure. The first type of such a model is the famous Boltzmann distribution. Subsequently, the Ising model \citep{ising_beitrag_1925} was developed for studying the magnetic dipole moments of atomic spins.

Through the GPP framework, we conduct a rigorous and integrated analysis of the spatial distributions of objects in the star formation complexes of the nearby galaxy M33 while simultaneously accounting for inhomogeneity effect and interpoint interaction. We adopt a hierarchical model structure to capture the natural formation hierarchy between GMCs and YSCCs. We also analyse how properties of the objects affect their spatial distributions without any information loss. 

The paper is organised as follows: section \ref{sec:methods} provides an introduction to point process theory, the GPP model and its meaningful construction. Section \ref{sec:models} contains a brief introduction to the data and the GPP models constructed as well as preliminary validation tests of the models. Section \ref{sec:data} includes the results of fitted models as well as model criticism. Section \ref{sec:discussion} provides a discussion on the comparison of our results to previous studies and potential physical implications from fitted models. Section \ref{sec:summary} gives the summary.

\section{Background and Methods} \label{sec:methods}

\subsection{Spatial Point Processes} \label{subsec:spp}

Spatial point process modelling concerns the study of the locations of the occurrence of random objects or events \citep{daley_introduction_2003, cressie_statistics_2011, baddeley_spatial_2007, baddeley_spatial_2015}. In this section, we only provide a brief introduction to the theory of point processes and the construction of Gibbs point processes. Readers interested in an introductory yet complete overview of the topic can refer to \cite{baddeley_spatial_2007, baddeley_spatial_2015}. A  rigorous mathematical treatment on the subject through measure theoretic probability is given by \cite{daley_introduction_2003, daley_introduction_2008}.

Given an observation window $S \subset \mathbb{R}^d$, where $d$ is 2 in our case, a point process $\mathbf{X}$ is a simple (no two points are coincidental) counting process, where its realisation/configuration is $\mathbf{x} = \{x_1, x_2, ..., x_n\} \subset S$. $x_i$ denotes the coordinate of the $i$-th point in $\mathbf{x}$. $n(\mathbf{X})$, the number of points in $\mathbf{X}$, is a random variable taking non-negative integer values. We only concern ourselves with the case $n(\mathbf{X}) < \infty$. We then also call $\mathbf{X}$ a finite point process.

The most fundamental spatial point process is the Poisson point process (PPP) which represents complete spatial randomness. The analysis of all other point patterns is taken with respect to PPP. A PPP in $S$ is characterised solely by an intensity (number density) function $\lambda(s) \geq 0$ satisfying the condition $\int_A \lambda(s)ds < \infty$ for any $A \subset S$. It is important to note the difference between a probability density function and the intensity function in that the latter specifies on average how many points there are in a given region. Although related, the two concepts are fundamentally different. A PPP with intensity $\lambda(s)$ will on average have $\int_A\lambda(s)ds$ points in any region $A \subset S$. However, a probability density $p(s)$ defined on $S$ means the fraction of points occurring in $A$ out of all points in $S$ is $\int_A p(s)ds$. If $\lambda(s) \equiv \lambda$, i.e. the intensity is constant in $S$, then the point process is a homogeneous PPP (hPPP). A hPPP with $\lambda = 1$ is called a unit-rate Poisson point process.

A PPP, $\mathbf{X}$, on $S$ with intensity $\lambda(s)$ is equipped with a probability density function \citep{baddeley_spatial_2007, daley_introduction_2003, baddeley_spatial_2015}:
\begin{equation}
\label{density_PPP}
    f(\mathbf{x}) = \exp\left(|S| - \int_{S}\lambda(s)ds\right)\prod_{i=1}^{n(\mathbf{x})}\lambda(x_i),
\end{equation}
where $|S|$ denotes the area of $S$.

A generic point process $\mathbf{X}$ does not have a probability density function on its own \citep{baddeley_spatial_2007} and Equation \ref{density_PPP} in fact is a probability density function with respect to a unit-rate PPP. In this sense, a unit-rate PPP serves as a reference point for all other point processes to be defined. If a point process $\mathbf{X}$ satisfies certain necessary conditions \citep{daley_introduction_2003, baddeley_spatial_2007}, it will then be equipped with a probability density function $f(\mathbf{x})$ with respect to a unit-rate PPP and the following equation holds:
\begin{multline}
\label{density_PP}
        \mathbb{P}(\mathbf{X} \in \mathcal{F}) = \sum_{n = 0}^\infty\frac{e^{-|S|}}{n!}\int_S \dots \int_S \mathbf{1}\left[\{x_1, \dots, x_n\} \in \mathcal{F}\right] \\
        \times f\left(\{x_1, \dots, x_n\}\right)dx_1\dots dx_2.
\end{multline}
Equation \ref{density_PP} represents the probability that $\mathbf{X}$ will have configuration $\mathbf{x} = \{x_1, x_2, ..., x_n\}$ such that $\mathbf{x} \in \mathcal{F}$ with respect to a unit-rate PPP, where $\mathcal{F}$ here is a set of possible configurations.

\subsection{Intensity Measures of Point Processes} \label{subsec: intensity}

Intensity measures of a point process are fundamental in characterising the structure within the process. As noted in section \ref{subsec:spp}, a PPP is defined by an intensity function $\lambda(s)$. This  corresponds to the first-order intensity measure $\mu(A)$ of a more general point process $\mathbf{X}$ which is defined as
\begin{equation}
    \mu(A) = \int_A \lambda(s)ds,
\end{equation}
for any $A \subset S$. Similar to the case of PPP, $\mu(A)$ specifies how many points there are within region $A$ on average. For point processes other than PPP, there are also higher-order intensity measures. Higher-order intensity measures are not considered for PPP since they are identically one, representing the idea of independence of occurrence between points. The most important higher-order intensity measure is the second-order intensity measure which quantifies the number of pairs of points in any region. Let $B \subset \mathbb{R}^d \times \mathbb{R}^d$ be a bivariate product space and let $\alpha^{(2)}(B)$ be the number of distinct pairs of points from $\mathbf{X}$ within $B$, then the second-order intensity measure, $\lambda^{(2)}(x,y) \geq 0$, is defined as \citep{moller_statistical_2003}
\begin{equation}
    \alpha^{(2)}(B) = \int_B\lambda^{(2)}(x,y)dxdy.
\end{equation}
If $\lambda^{(2)}(x,y)$ is well-defined, then we can subsequently define the PCF $g(x,y)$ as \citep{moller_statistical_2003}
\begin{equation}
\label{pcf}
    g(x,y) = \frac{\lambda^{(2)}(x,y)}{\lambda(x)\lambda(y)}.
\end{equation}
If we assume further that the point process is second-order stationary, we then have
\begin{equation}
    g(x,y) = \rho(||x-y||),
\end{equation}
for some non-negative function $\rho$ and $||\cdot||$ being a metric such as the Euclidean distance. 

If $g(x,y) \equiv 1$, then $\mathbf{X}$ corresponds to the behaviour of a PPP. If $g(x,y) > 1$, then it means the point pattern is clustered/aggregated compared to a PPP at position $x$ and $y$. If $g(x,y) < 1$, then it represents a repulsive/inhibitive pattern compared to a PPP.

From Equation \ref{pcf}, it is clear the PCF depends on the first-order intensity $\lambda(s)$. This is a crucial mathematical demonstration of how inhomogeneity not being accounted for properly can lead to problematic conclusions from the empirical PCF/2PCF. The empirical PCF is generally different from the theoretical PCF given by Equation \ref{pcf} in that the empirical PCF is obtained by assuming the underlying point process is homogeneous, i.e., $\lambda(s)$ is constant. However, this is rarely the case for point patterns comprised of SCs due to their highly inhomogeneous distribution across the galaxy disc, e.g., decreasing projected density (intensity) with galactocentric distance.

\subsection{Gibbs Point Process}

Assuming that a point process $\mathbf{X}$ satisfies $n(\mathbf{X}) < \infty$, then it is a finite Gibbs point process if it has probability density $f(\cdot)$ in the sense of equation \ref{density_PP}, such that $f(\cdot)$ can be written as \citep{baddeley_spatial_2007, baddeley_spatial_2015}:
\begin{equation}
    f(\mathbf{x}) = \exp\left(V_0 + \sum_{x \in \mathbf{x}}V_1(x)
    +\sum_{\{x,y\}\subset \mathbf{x}}V_2(x,y) + \dots\right),
\end{equation}
where $x, y$ are distinct points in $\mathbf{x}$. $V_k$ is called the $k-$th order potential. Potentials of order $>2$ are generally termed  higher-order potentials.

From equation \ref{density_PPP}, for a hPPP with intensity $\lambda$, $V_0 = |S|(1 - \lambda)$, $\exp(V_1(x)) = \lambda, \ \forall x \in \mathbf{x}$, and $V_k = 0, \forall k \geq 2$. In fact, for any PPP, the second-order and all higher-order potentials vanish. This corresponds to the feature of PPP that there exists no interpoint interaction between points. The first-order potential is the ``fertility" effect mentioned earlier and it is used to characterise the external inhomogeneous effects, such as covariates, on the intensity of the point process, while potentials with order $k \geq 2$ together characterise the dependence structure within a point process. 

\subsubsection{Pairwise-Interaction Process} \label{subsubsec: pwp}

In many practical applications, higher-order potentials are set to zero for simplicity and the dependence structure within the point process is  solely characterised by the second-order potential. In this case, the GPP model reduces to the so-called pairwise-interaction process \citep{baddeley_spatial_2007, baddeley_spatial_2015}, and it is this model framework that we will adopt in this paper.

A pairwise-interaction process will then have probability density function 
\begin{equation}
\label{density_PW}
    f(\mathbf{x}) = \exp\left(V_0 + \sum_{x \in \mathbf{x}}V_1(x)
    +\sum_{\{x,y\}\subset \mathbf{x}}V_2(x,y)\right)
\end{equation}
with respect to a unit-rate PPP. Equation \ref{density_PW} is also called the canonical form. After reparameterization, the canonical form can also be written as 
\begin{equation}
    \label{repar}
    f(\mathbf{x}) = \alpha \prod_{x \in \mathbf{x}}\lambda(x) \prod_{\{x,y\}\subset \mathbf{x}} \phi(x,y),
\end{equation}
where $\alpha = \exp(V_0), \lambda(x) = \exp(V_1(x))$, and $\phi(x,y) = \exp(V_2(x,y))$.

The main task of modelling is then to specify the structure of $\lambda(x)$ and $\phi(x,y)$ based on the behaviour of the data. This means that for most models, $V_0$ or $\alpha$ is unknown, i.e., the part of probability density that we can work with is the unnormalised density $h(\mathbf{x}) = f(\mathbf{x})/\alpha$.

As mentioned, $\phi(x,y)$  characterises the interpoint interaction. However, $\phi$ also has close ties with the 2PCF/PCF. If we assume that the first-order potential or $\lambda(x)$ is constant, i.e., the point process is homogeneous, then $\phi(x,y)$ is also a parametric representation of the 2PCF/PCF of the point pattern \citep{goldstein_attraction-repulsion_2014}. In this case, one can simply obtain a fitted empirical PCF and choose a parametric form of $\phi(x,y)$ accordingly. However, this does not equate to saying that the PCF is the same as $\phi(x,y)$, only that $\phi(x,y)$ will closely resemble the structure of the PCF. Notably, the empirical PCF can no longer be used to inform our decision in choosing $\phi(x,y)$ since the two are no longer quantitative counterparts due to the first-order inhomogeneity.

In practice, it is generally assumed that the point pattern analysed is second-order stationary unless there is clear evidence to suggest otherwise. Similar to the case of PCF, second-order stationarity implies that
\begin{equation}
\label{second-order}
    V_2(x,y) = h(||x-y||), \forall x,y \in \mathbf{x}
\end{equation}
for some function $h(\cdot)$. $||\cdot||$ is the standard Euclidean metric. In this paper, we assume our point pattern is stationary.

There are infinitely many ways to construct GPP models. However, well-defined construction is not arbitrary as the constructed model needs to satisfy certain stability criteria, namely local stability and Ruelle stability \citep{moller_statistical_2003, baddeley_spatial_2007, baddeley_spatial_2015}. These criteria are needed to ensure the existence of the probability density function of GPP models. Interested readers can refer to the aforementioned references for details or Appendix~\ref{sec:stability} for a self-contained introduction. 

For conducting inference of GPP models, one needs to be able to simulate a point pattern based on a specified GPP model. This is generally done through the Birth-Death Metropolis-Hastings algorithm \citep{geyer_simulation_1994}. For inference algorithms, there are many different approaches one could choose from based on the inference paradigms (maximum likelihood estimation/Bayesian inference). We provide the details of these algorithms in Appendix~\ref{algorithms}.

\subsubsection{Bivariate and Hierarchical Processes} \label{subsubsec: multi}

If the point patterns considered consist of multiple types of points, such GMCs and YSCCs in our case, and we are interested in how they interact/correlate with each other, then a bivariate point process should then be considered. Here we give a brief outline for the bivariate point process under the framework of GPP. For simplicity, we only restrict our scope under the assumption of pairwise-interaction.

Suppose two point processes, $\mathbf{X}_A$ and $\mathbf{X}_B$, form a bivariate Gibbs point process \citep{isham_multitype_1984}, assuming the functional form of equation \ref{repar}, then it has the following joint probability density function
\begin{multline}
     \label{multi-type}
    f(\mathbf{x}_A, \mathbf{x}_B) = \alpha\lambda_A^{n(\mathbf{x}_A)}\lambda_B^{n(\mathbf{x}_B)}\phi_A(\mathbf{x}_A)\phi_B(\mathbf{x}_B)\\
    \times\phi_{AB}(\mathbf{x}_A,\mathbf{x}_B).
\end{multline}
Here we assume both point processes are homogeneous and the notations are abbreviated to save space. $\lambda_A, \lambda_B$ control the first-order potential, $\phi_A(\mathbf{x}_A), \phi_B(\mathbf{x}_B)$ characterise the intra-type interaction in $\mathbf{x}_A, \mathbf{x}_B$ respectively. The extra term $\phi_{AB}(\mathbf{x}_A,\mathbf{x}_B)$ denotes the inter-type interaction/correlation between the points of $\mathbf{x}_A, \mathbf{x}_B$. Similar to the relationship between the empirical 2PCF/PCF of $\mathbf{x}_A$ and $\phi_A(\mathbf{x}_A)$, $\phi_{AB}(\mathbf{x}_A,\mathbf{x}_B)$ also has an empirical counterpart in that it represents the cross-type 2PCF/PCF \citep{baddeley_spatial_2015} between $\mathbf{x}_A$ and $\mathbf{x}_B$. The cross-type PCF is a generalisation of the PCF in that it measures the ratio of the probability of observing a point in the first type at $r$ distance away from a point in the second type to that of a case where the two are uncorrelated.

If further information is available that there exists a form of hierarchy between two types of points, i.e., one type takes precedence over another, then it is more appropriate to consider a hierarchical structure between the two processes through conditional probability density.

The bivariate hierarchical Gibbs point process was first considered by \cite{hogmander_multitype_1999} to analyse the point patterns of the nests of two species of ants. However, one species of ants exhibit ecological dominance over the other. In this scenario where there is a natural order or asymmetry between types of points, it is no longer appropriate to formulate the model through the bivariate point process. \cite{hogmander_multitype_1999} then proposed a hierarchical point process through conditional probability argument. First, they define the process that takes precedence, $\mathbf{X}_A$, as a high-level univariate GPP:
\begin{equation}
    f(\mathbf{x}_A) = \alpha_A\lambda_A^{n(\mathbf{x}_A)}\phi(\mathbf{x}_A).
\end{equation}
Then they define the low-level process, $\mathbf{X}_B$, as a conditional process given the configuration of $\mathbf{x}_A$:
\begin{equation}
\label{hierarchy}
    f_{\mathbf{x}_A}(\mathbf{x}_B) = \alpha_B(\mathbf{x}_A)\lambda_B^{n(\mathbf{x}_B)}\phi_B(\mathbf{x}_B)\phi_{AB}(\mathbf{x}_A, \mathbf{x}_B).
\end{equation}
The difference between the formulation of \ref{multi-type} and \ref{hierarchy} is subtle but the philosophy behind the model construction is fundamentally different. This also leads to different approaches for model fitting. To fit a bivariate point process, $\mathbf{x}_A, \mathbf{x}_B$ are considered simultaneously. For the hierarchical point process, one has to fit the model for the high-level process first, then treat $\mathbf{x}_A$ as given when fitting the model for $\mathbf{x}_B$. However, if $\mathbf{x}_A$ is not the main study interest, fitting a single model on $\mathbf{x}_B$ is also possible by simply treating the pattern of $\mathbf{x}_A$ as a fixed underlying structure. Importantly, the hierarchical structure does not mean $\mathbf{x}_A$ does not depend on $\mathbf{x}_B$, only that the dependence of $\mathbf{x}_A$ on $\mathbf{x}_B$ is not specified explicitly \citep{baddeley_spatial_2015}.

We consider this hierarchical structure due to the natural formation hierarchy present between GMCs and YSCCs. Since there is a reasonable amount of evidence to suggest that YSCCs are born out of GMCs \citep{carlberg_magnetic_1990, mclaughlin_formation_1996, grasha_spatial_2019}, we assume the formation of GMCs takes precedence before YSCCs. 


\section{Data and Model Construction} \label{sec:models}

\begin{figure*}
    \centering
    \subfloat[CO filament and GMCs]{%
    \includegraphics[width = 100mm]{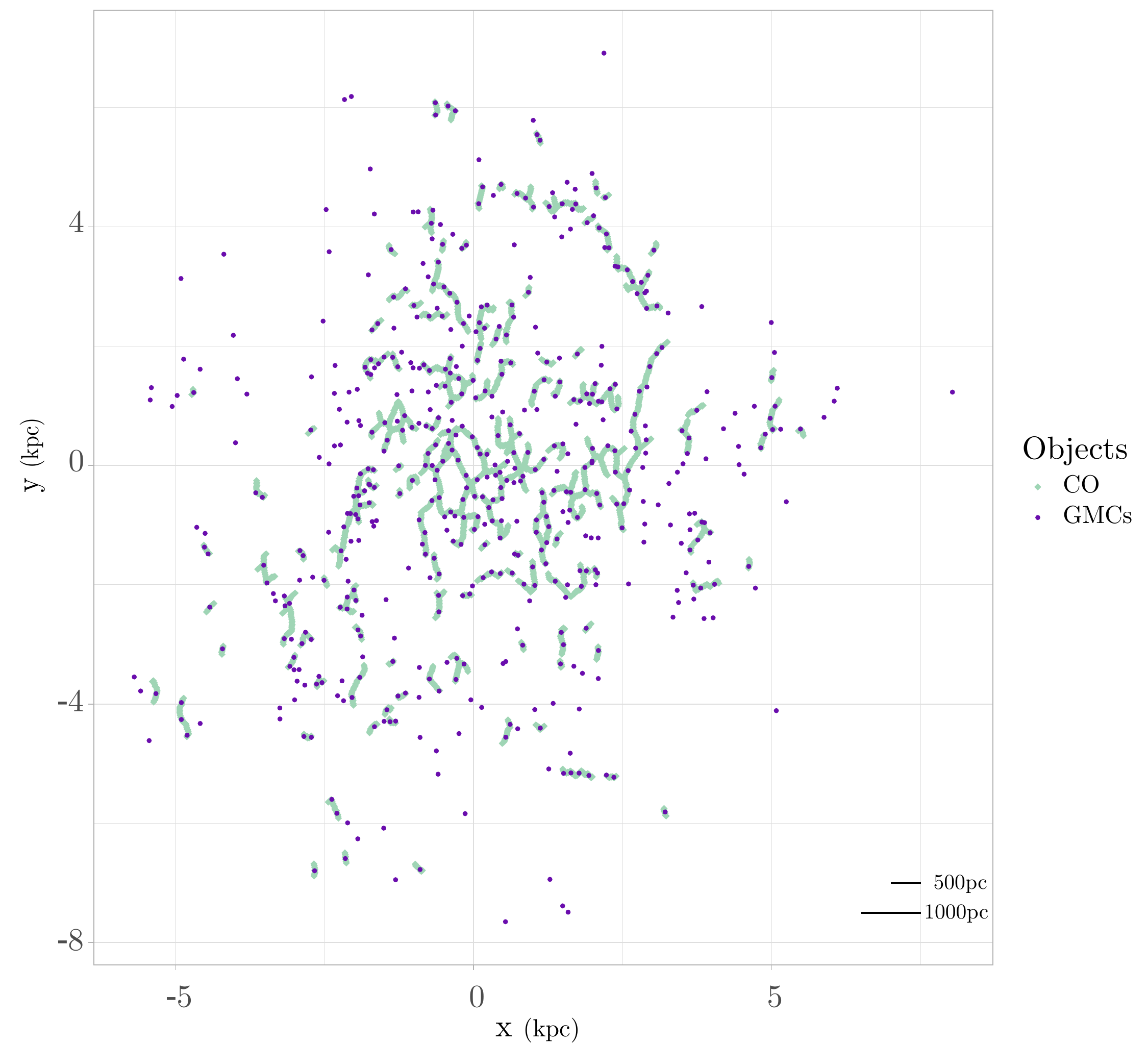}%
    }
    \hspace{10mm}
    \subfloat[GMCs and YSCCs]{%
    \includegraphics[width = 100mm]{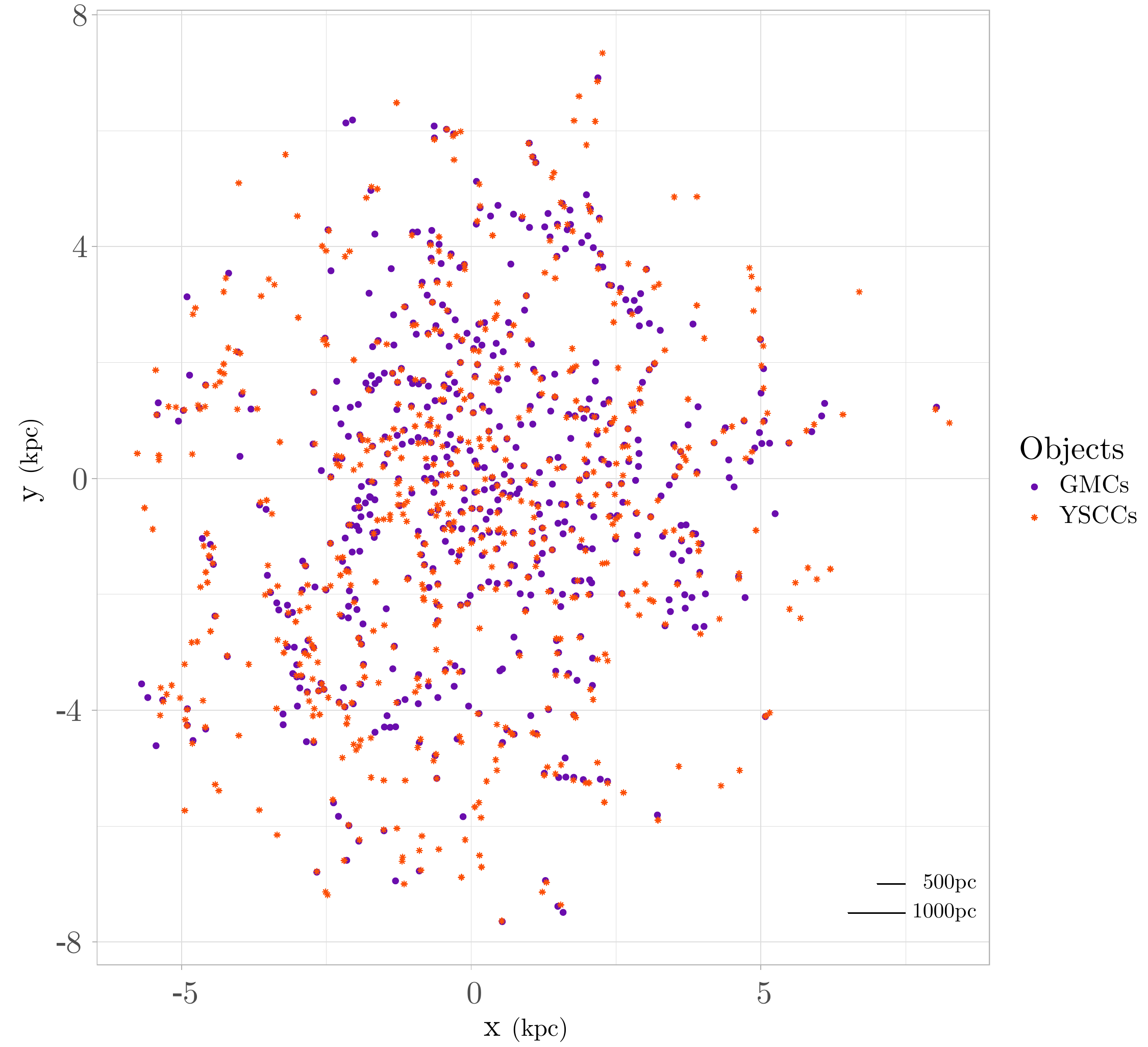}%
}
    \caption{(a) Overlay of CO filament and GMCs in M33; (b) Overlay plot of GMCs and YSCCs in M33.}
    \label{fig:CGS}
\end{figure*}

We choose M33 for our analysis since it is one of the few low-inclination galaxies with a relatively complete catalog of GMCs. Three sets of data are used in the analysis: maps of the CO filament structure, observations of GMCs, and observations of YSCCs. We include the CO filament structure since we want to investigate how it can potentially affect the distribution of YSCCs. The IRAM 30-m observations of CO(2-1) emission were published by \citet{druard_iram_2014}. The CO filamentary structure was obtained\footnote{Kindly provided by Eric Koch and Erik Rosolowsky} using the method described in \cite{koch_filament_2015}. The GMCs are also identified by \cite{corbelli_molecules_2017} using the IRAM 30-m observations of CO(2-1) emission in \cite{druard_iram_2014} and the YSCCs are identified using  Spitzer 24-$\mu$m observations, published by \cite{sharma_population_2011, corbelli_molecules_2017}. The data consist of the positions, galactocentric distance, effective radius, velocity dispersion, gas mass, and virial mass of 566 identified GMCs and the positions, size, and (incomplete) estimates of age and mass of 630 identified YSCCs. We consider both confirmed and candidate young stellar clusters (YSCs) since there are only around 400 identified YSCs (with estimation of mass and age). Furthermore, the 630 candidate YSCs are what was analysed in \cite{corbelli_molecules_2017} and it is appropriate for us to also use the candidates catalog for drawing comparison.

To account for the inclination of M33, we first carry out a coordinate transform of the data from RA/DEC to the 2D projected Cartesian coordinates, assuming the distance to M33 is $D=$840 kpc \citep{bonanos_first_2006, magrini_planetary_2009}, an inclination of $i = 53^\circ$ \citep{magrini_planetary_2009} and the position angle $\theta = 22^\circ$ \citep{magrini_planetary_2009}.

Figure \ref{fig:CGS} show plots of the CO filamentary structure, GMCs and YSCCs in the 2D projected Cartesian coordinates. Simply from visual inspection, the GMCs and YSCCs are highly correlated.

We construct the model for YSCCs through the hierarchical point process framework by treating the point pattern of GMCs as given, i.e., it is regarded as fixed. Denoting the point pattern of GMCs as $\mathbf{x}_G$ and the point pattern of YSCCs as $\mathbf{x}_S$, we follow the general form of the model given by equation \ref{hierarchy} and write out the likelihood function:
\begin{multline}
    f_{\mathbf{x}_{G}}(\mathbf{x}_{S}; \ \boldsymbol{\theta}_{S}) = \mathcal{L}(\boldsymbol{\theta}_{S}|\mathbf{x}_{S}; \mathbf{x}_{G})= \\
    \alpha_{S}(\mathbf{x}_{G})\prod_{j = 1}^{n(\mathbf{x}_{S})}\lambda_{S}(x_{j,S})\phi_{S}(\mathbf{x}_{S})\prod_{i = 1}^{n(\mathbf{x}_G)}\prod_{j = 1}^{n(\mathbf{x}_S)}\phi_{GS}(x_{i,G}, x_{j,S}).
\end{multline}
 Here, $\boldsymbol{\theta}_{S}$ is the vector of model parameters. $\lambda_{S}(x_{j,S})$ is the first-order potential at the location of the $j$-th YSCC, $\phi_S(\mathbf{x}_S)$ is the second-order potential for YSCCs, and $\phi_{GS}(x_{i,G}, x_{j,S})$ is the correlation between the $i$-th GMC and the $j$-th YSCC. $\alpha_{S}(\mathbf{x}_{G})$ is the unknown normalising constant dependent on the parameters and $\mathbf{x}_G$. We now give the parametric structure for each term.
 
\begin{figure}
    \centering
    \includegraphics[width = 80mm]{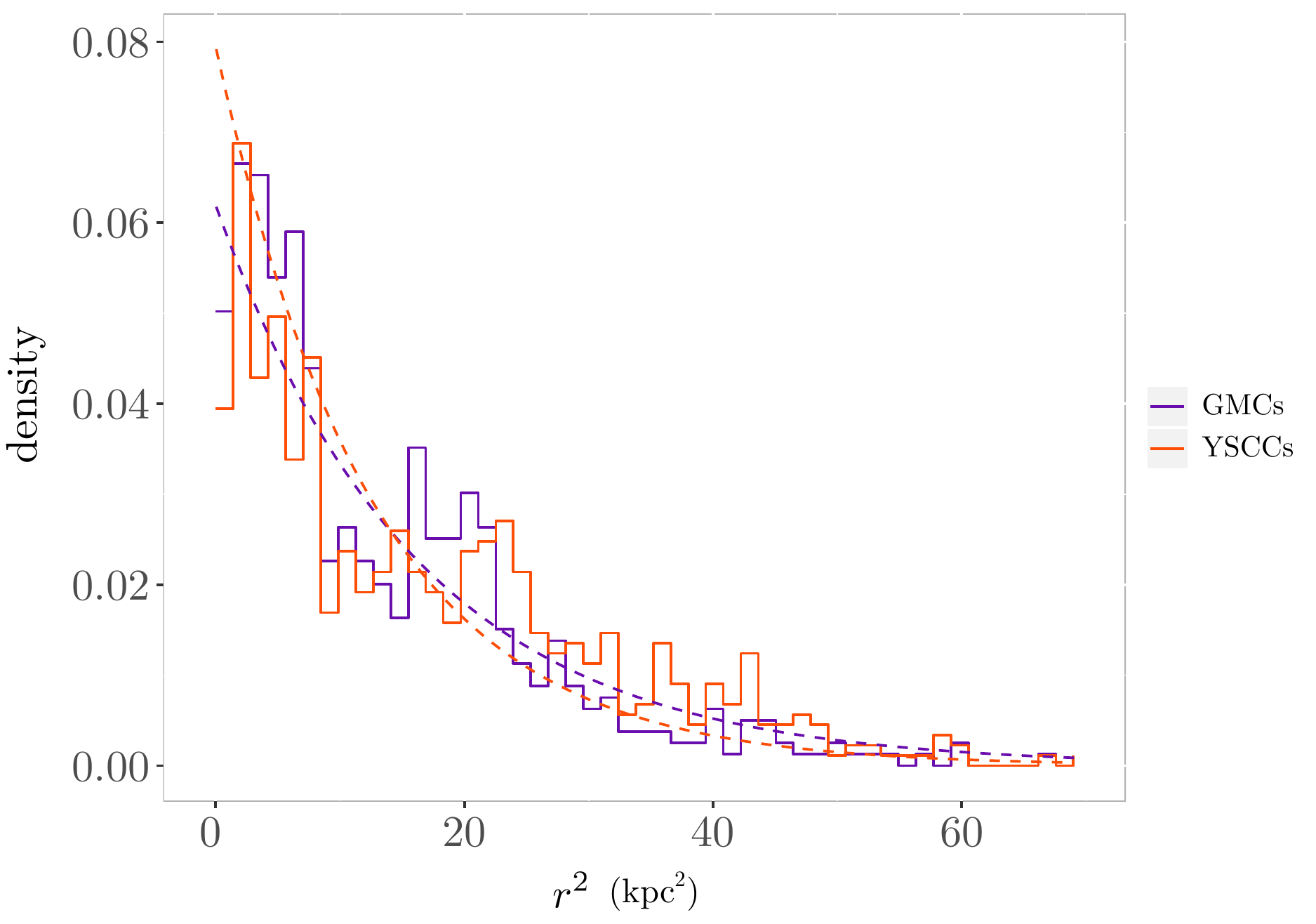}
    \caption{Exponential fit of the squared galactocentric distance.}
    \label{hist_gmc_sc_d}
\end{figure}
 
For the large scale distribution of YSCCs, we consider it to be a function of the galactocentric distance. Since the general large scale distributions of GMCs and YSCCs are both approximately normal centred around the galaxy centre, their squared galactocentric distance will approximately follow an exponential distribution as shown in Figure \ref{hist_gmc_sc_d}. The overlapping large scale distribution of GMCs and YSCCs will be a lurking variable that can undermine the investigation of the actual correlation structure between GMCs and YSCCs. Therefore, this distribution will be accounted for in the first-order potential term as a large scale spatial trend:
 \begin{equation}
 \label{first-order}
     \lambda_S(x_{j,S}) = \exp\left(P_2(x_{j,S};\boldsymbol{p})\right)
 \end{equation}
 where $P_2(x_{j,S};\boldsymbol{p})$ is a second-order polynomial function of  the distance from the $j^{\text{th}}$ YSCC to the galactic centre. To make the model as simple as possible, we assume the following form for $P_2$:
 \begin{equation}
 \label{large-scale}
      P_2(x_S; \rho, R_{s,c}) = -\left(\frac{r_{s,c}}{R_{s,c}}\right)^2 + \rho,
 \end{equation}
 where $r_{s,c}$ is the distance from YSCC $x_S$ to the galaxy centre, $R_{s,c}$ is the characteristic scale of the distribution of YSCCs in the galaxy disc, and $\rho$ controls the large scale intensity at the centre of the galaxy.
 
For the correlation between the GMCs and YSCCs, we choose the following parametric form:
\begin{equation}
\label{correlation}
    \phi_{GS}(x_{i,G}, x_{j,S}) = \exp\left[\psi_i\left(1 + \frac{r_{ij}^2}{\sigma_{GS}^2}\right)^{-\frac{5}{2}}\right].
\end{equation}
In this model, $\psi_i$ controls the correlation strength between the $i$-th GMC and all YSCCs. The greater the value of $\psi_i$, the greater the correlation between GMCs and YSCCs. $r_{ij}$ is the distance between the $i^{\text{th}}$ GMC and the $j^{\text{th}}$ YSC. $\sigma_{GS}$ is a characteristic scale parameter controlling the correlation scale between GMCs and YSCCs. Notice that if $\psi_i = 0$, it then suggests that there is no correlation between GMCs and YSCCs. 

\begin{figure}
    \centering
    \includegraphics[width = 80mm]{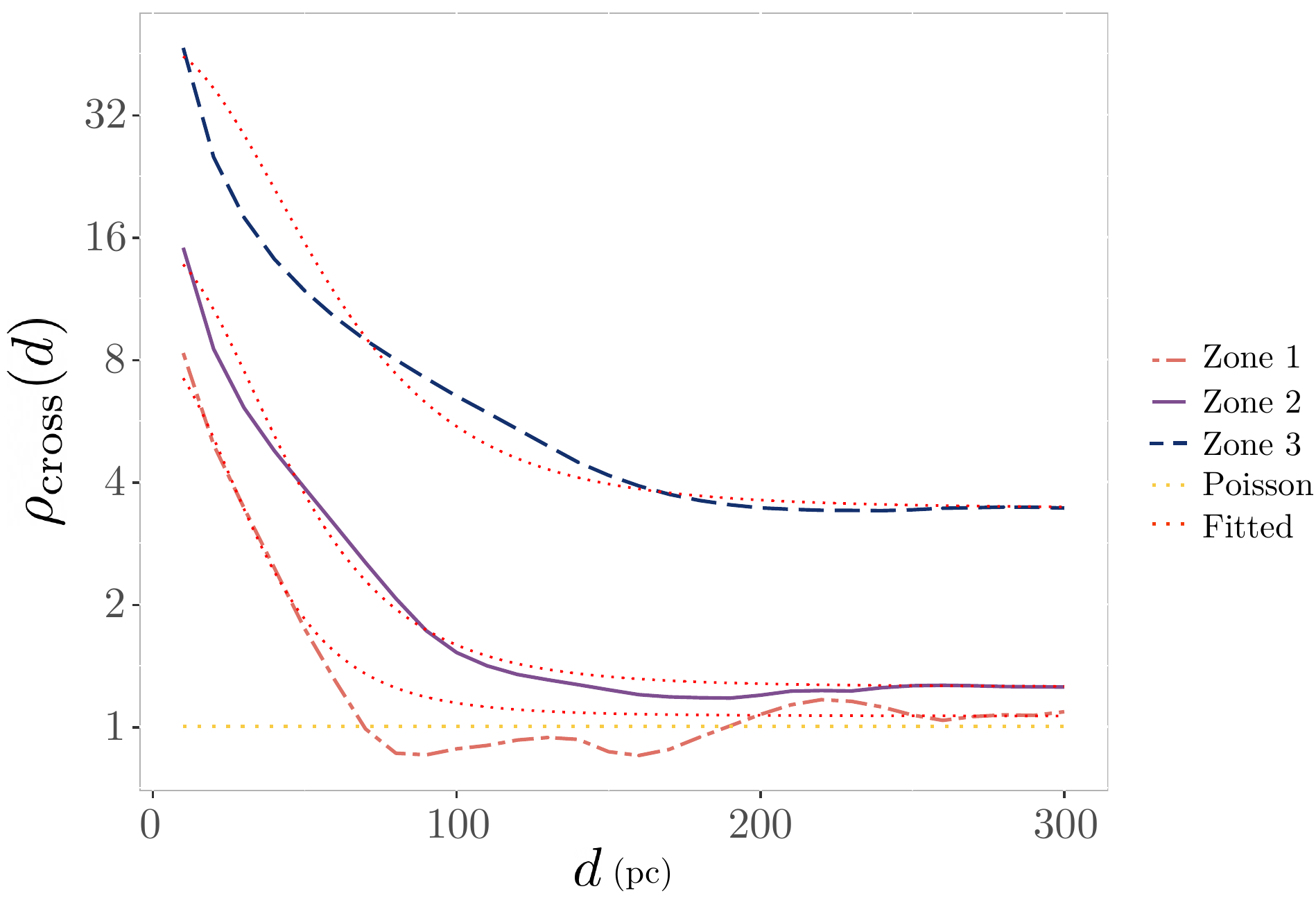}
    \caption{Cross-type PCF between GMCs and YSCCs; zone 1: $D < 1.5$ kpc; zone 2: $1.5$ kpc $\leq D < 4$ kpc; zone 3: $D \geq 4$ kpc where $D$ is the galactocentric distance; $d$ is the distance between GMC and YSCC}
    \label{cpcf}
\end{figure}

We assume the distribution of YSCCs around each YSCCs follows a Plummer (5,2) power law \citep{plummer_problem_1911, dejonghe_completely_1987} for simplicity. In theory, we can also set the power of the correlation as a free parameter to be fitted. However, doing this will drive up the number of parameters and increases computational complexity. Moreover, a preliminary analysis on the cross-type 2PCF/PCF (in log-scale) between GMCs to YSCCs shows a similar power law shape as indicated in Figure \ref{cpcf}. We do this in the same fashion as in \cite{corbelli_molecules_2017} by dividing the the galaxy disc into three zones based on the galactocentric distance (zone 1: $D < 1.5$ kpc; zone 2: $1.5$ kpc $\leq D < 4$ kpc; zone 3: $D \geq 4$ kpc). We also fitted a modified Plummer (5,2) power law for each zone based on their cross-type PCF denoted by red dotted lines in Figure \ref{cpcf}. We see that for zone 1 and zone 2, the power law structure is indeed quite similar to a (5,2) power law. However, there is significant difference for zone 3. This is because the point pattern in zone 3 still exhibits inhomogeneous behavior as the cross-type PCF does not drop to unity when distance increases. Therefore, the resulting behavior of cross-type PCF can be quite different from a power law structure. However, we can conclude that the Plummer (5,2) structure is close enough to the actual correlation structure. The correlation scale does increase from zone 1 to zone 3 but the difference is rather small. Furthermore, fitting a varying correlation scale parameter is rather computationally problematic.

\begin{figure}
    \centering
    \includegraphics[width = 80mm]{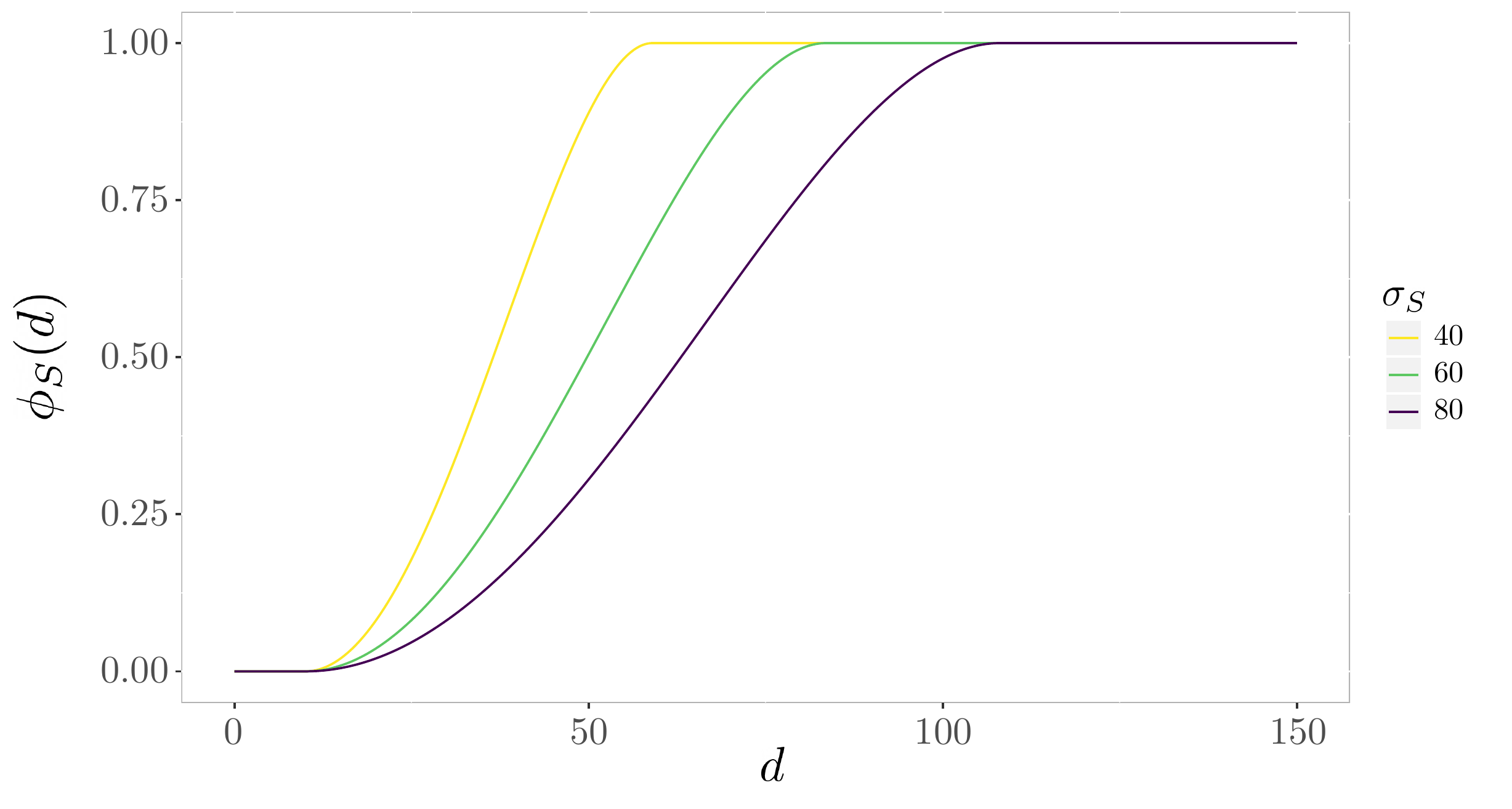}
    \caption{Plot of $\phi_S(d)$ with different $\sigma_S$ values. $\phi_S(d)$ is the second-order intensity value as a function of interpoint distance $d$ between two YSCCs; $\sigma_S$ is the characteristic scale of inhibitive structure within YSCCs}
    \label{phi_S}
\end{figure}

Since we are considering all possible pairings between GMCs and YSCCs, choosing the formulation described above circumvents the problems in rudimentary analysis where YSCCs are assigned an associated GMC by nearest neighbour distance. This eliminates a potential bias introduced by wrongful nearest neighbour assignment.
 
The reason we emphasise a different correlation strength parameter $\psi_i$ for each GMC is to facilitate an accurate and quantitative investigation of how the properties of the GMCs affect the correlation between GMCs and YSCCs. With available data from \cite{corbelli_molecules_2017}, there are three properties of GMCs of interest: (1) galactocentric distance of GMCs $D$, which is already shown in Figure \ref{cpcf} to have an effect on the correlation; (2) the log-mass of a GMC $\log_{10}(M/M_\odot)$; (3) the log-NN distance from a GMC to the CO filament $\log_{10}(d_{gc})$. We model their effects by a simple linear-regression-like structure:
\begin{equation}
    \psi_i = \theta_0 + \theta_{D}D_i + \theta_{M}\log_{10}(M_i/M_\odot) + \theta_{gc}\log_{10}(d_{i,gc}).
\end{equation}
When conducting model fitting, we normalise the properties' data with respect to the mean and standard deviation. This is for better comparison between the effects of different properties on the correlation strength. The interpretation of the parameters is similar to the case in a linear model. $\theta_0$ is the baseline correlation strength between GMCs and YSCCs. It is also the average correlation between GMCs and YSCCs if we standardise the data of each property. Other parameters are similar to the slope parameters in a linear model. For example, if $\theta_D > 0$, then it means $D$ has a positive effect on the correlation between GMCs and YSCCs, i.e., GMC which has a greater galactocentric distance is more correlated with YSCCs, and vice versa. If $\theta_D=0$, then $D$ does not have any effect on the correlation.
 
\begin{table*}
 \begin{center}
    \caption{Model Parameters}
    \label{tab:param}
    \begin{tabular}{l|c|c}
      \toprule 
      Parameters & Meaning & Domain\\
      \midrule 
      $R_{s,c}$ (kpc) & Characteristic scale of the large scale variation of YSCCs across the galaxy disc & $(0, \infty)$\\
      $\rho$ & Log-intensity of YSCCs at the centre of the galaxy & $\mathbb{R}$ \\
      $\theta_0$ & Baseline correlation strength between GMCs and YSCCs & $\mathbb{R}$\\
      $\theta_D$ & The effect of galactocentric distance of GMCs on the correlation strength between GMCs and YSCCs & $\mathbb{R}$\\
      $\theta_M $ & The effect of mass of GMCs on the correlation strength between GMCs and YSCCs & $\mathbb{R}$\\
      $\theta_{gc} $ & The effect of distance from GMCs to CO filament on the correlation strength between GMCs and YSCCs & $\mathbb{R}$ \\
      $\sigma_{GS}$ (pc) & Characteristic scale of correlation between GMCs and YSCCs & $(0, \infty)$\\
      $\sigma_{S}$ (pc) & Characteristic scale of inhibitive structure among YSCCs & $(0, \infty)$\\
      \bottomrule 
    \end{tabular}
  \end{center}
\end{table*}
 
For the second-order intensity, assuming stationarity, we employ the following model:
  \begin{multline}
      \phi_{S}(d_{kl}) = 
    \begin{dcases*}
        0,  \ 0 < d_{kl} \leq R_S,\\
        \frac{4}{3}\left(\frac{d_{kl} - R_S}{\sigma_S}\right)^2\\
        \times\left(1 - \left(\frac{d_{kl} - R_S}{\sqrt{3}\sigma_S}\right)^2\right),  R_S < d_{kl} \leq R_P, \\
        1, \ d_{kl} > R_P,
    \end{dcases*}
  \end{multline}
  
\begin{figure*}
    \centering
    \includegraphics[width = 185mm]{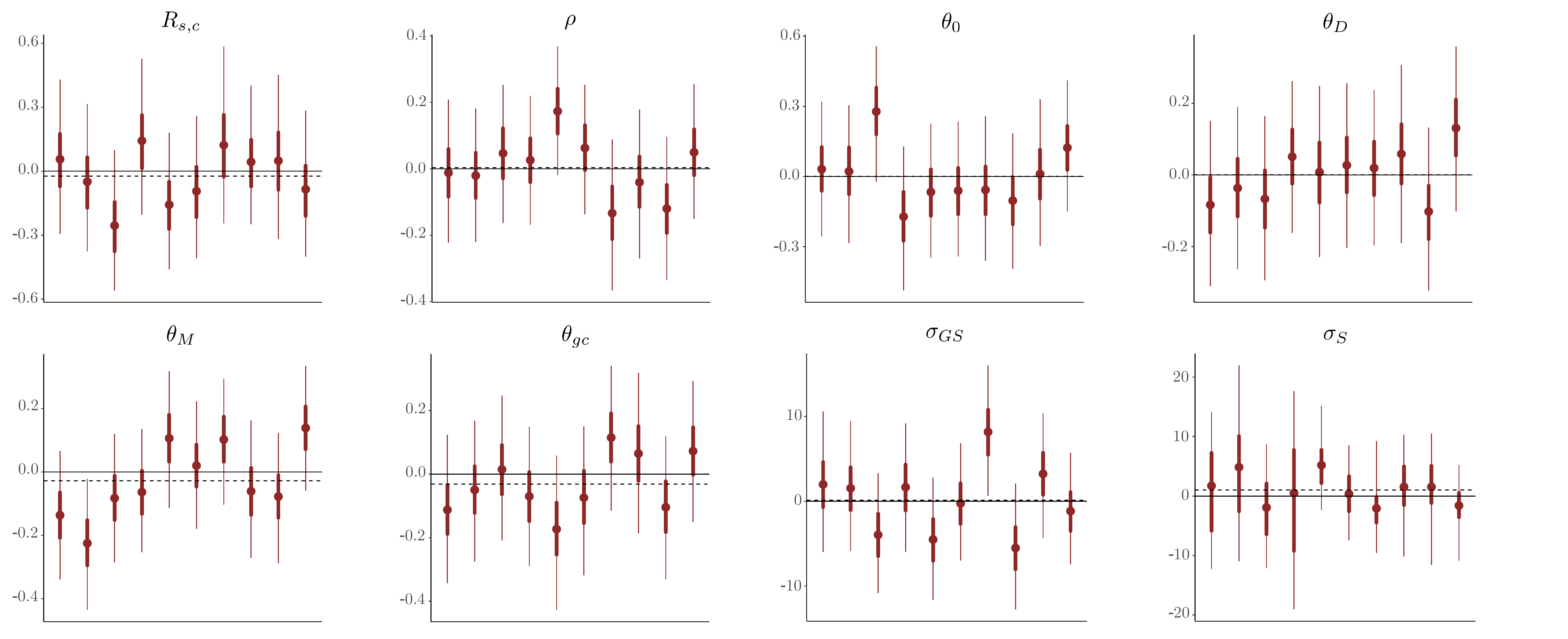}
    \caption[Plot of bias-adjusted posterior samples inferred from 10 simulated data sets under parameter configuration \{$R_{s,c} = 4.65$, $\rho = 0.5$, $\theta_0 = 4$, $\theta_D = 0.5$, $\theta_M = 0.5$, $\theta_{gc} = 0$, $\sigma_{GS} = 89$, $\sigma_S = 54$\}.]{Plot of bias-adjusted posterior samples inferred from 10 simulated data sets under parameter configuration\{$R_{s,c} = 4.65$, $\rho = 0.5$, $\theta_0 = 4$, $\theta_D = 0.5$, $\theta_M = 0.5$, $\theta_{gc} = 0$, $\sigma_{GS} = 89$, $\sigma_S = 54$\}. The thick red line segments denotes the 50\% credible intervals of bias while the thin red lines are the 95\% credible intervals. The red circles are the estimated posterior mean biases. The horizontal black solid lines are the reference baseline of zero bias. The dotted black lines are the average bias obtained from all posterior samples.}
    \label{fig:model_validate}
\end{figure*}
where $R_P = \sqrt{3/2}\sigma_S + R_S$. $d_{kl}$ is the distance between the $k$-th YSCC to the $l$-th YSCC. $\sigma_S$ is a characteristic scale that determines the range of inhibitive effect between two YSCCs. However, $R_p$ here is the actual parameter representing the inhibitive scale. Figure \ref{phi_S} shows the shape of $\phi_S(d)$ with different choices of $\sigma_S$.
We choose this formulation for the second-order structure since the empirical PCF can no longer be used to determine the actual second-order property for the YSCCs due to the obvious inhomogeneity of YSCCs distribution. 

The justifications for the choice of this form of second-order potential are the following: (a) it is easy to implement and has guaranteed numerical stability. Furthermore, the second-order potential is smooth and differentiable at all scales; (b) YSCCs all have physical sizes denoted by $R_S$. If two YSCCs are at the same location, they will eventually be identified as one YSCC, and as noted, we do not consider cases where there exist coincidental points; (c) at very short scales, the distribution of YSCCs should be inhibitive since there exists competition for the star formation fuel. Furthermore, the stellar feedback can blow away surrounding gas in the molecular clouds and regulate star formation rate \citep{grasha_spatial_2019, chevance_lifecycle_2019}. This is also demonstrated in the simulation by \cite{rogers_feedback_2013}. The stellar feedback and blowouts of SCs on their surrounding molecular gas in fact corresponds to a form of ``competition'' for star forming resources. This means that in a small and compact region, it is unlikely for two YSCCs to exist.  Although it might happen that two YSCCs can become gravitationally bound with each other, the probability of observing this should be very small. However, we need to point out that for pairwise distance within $R_P$, it does not mean there cannot be more than one YSCC, rather it only means that the chance of finding two YSCCs within this distance is less than that of a Poisson process. At larger scales, there might still be interpoint interaction among YSCCs, however, we cannot use the empirical PCF/2PCF to inform ourselves as to what type of behaviour YSCCs exhibit among themselves. Therefore, we set the pairwise interaction to one, i.e., we assume that the YSCCs do not interact with each other at larger scale. We can then infer their actual behaviour at larger scales from model criticism since any discrepancy between the data and model can be easily interpreted. This is because the model, as a reference, is a Poisson process at the greater range.
 
Table \ref{tab:param} gives an overview of the model parameters for  reference.

Before we conduct data analysis for the real data, we first need to confirm that the constructed models are well-behaved enough so that the inference algorithms can recover the model parameters. We consider ten sets of simulated data from the Birth-Death Metropolis-Hastings (BDMH; see Appendix~\ref{simulation}) algorithm and conduct inference on the simulated data. The parameter set chosen is \{$R_{s,c} = 4.65$ kpc, $\rho = 0.5$, $\theta_0 = 4$, $\theta_D = 0.5$, $\theta_M = 0.5$, $\theta_{gc} = 0$, $\sigma_{GS} = 89$ pc, $\sigma_S = 54$ pc\}. The results are shown in Figure \ref{fig:model_validate}. The figure shows a bias-adjusted posterior distribution where the true parameter values are subtracted from each posterior sample. The thick red line segments denotes the 50\% credible intervals of bias against the true parameters obtained through the posterior distributions while the thin red lines are the 95\% credible intervals. The red circles are the estimated posterior mean biases. The dotted black lines are the average bias obtained from all posterior samples. 

 \begin{table*}
  \begin{center}
    \caption{Crude estimate of GMC-SC model parameters}
    \label{tab:gmc_sc_crude}
    \begin{tabular}{c|c|c|c|c|c|c|c}
      \toprule 
      $\log(R_{s,c}^0)$ (kpc) & $\rho^0$ & $\theta_0^0$ & $\theta_D^0$ & $\theta_M^0$ & $\theta_{gc}^0$ & $\log(\sigma_{GS}^0)$ (pc) & $\log(\sigma_{S}^0)$ (pc)  \\
      \midrule 
       log(5) & 0.8 & 4 & 0 & 0 & 0 & log(76) & log(100) \\
      \bottomrule 
    \end{tabular}
  \end{center}
\end{table*}

\begin{figure*}
    \centering
    \includegraphics[width = 170mm]{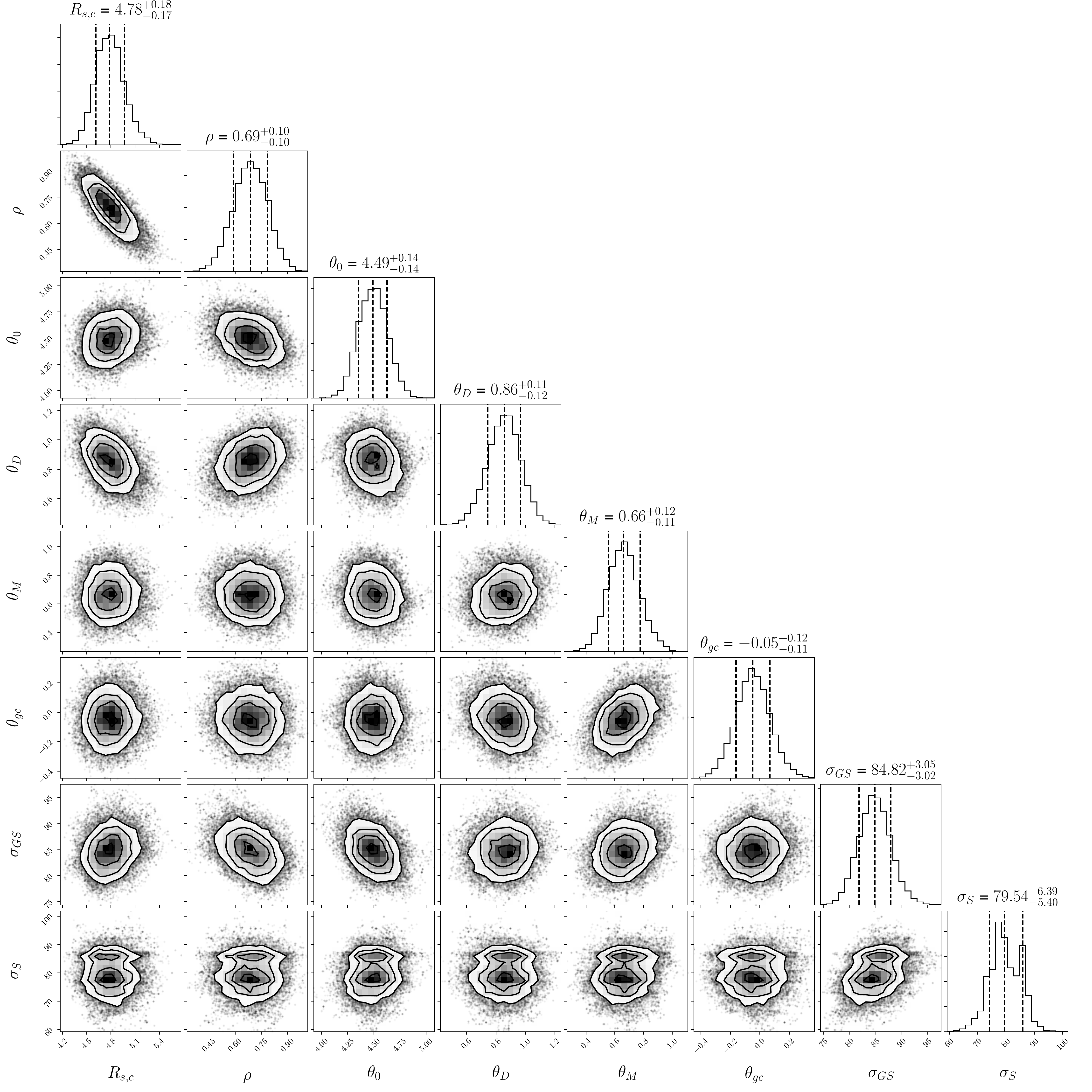}
    \caption{Corner plot of the posterior distribution of the model with estimated posterior mean and $68\%$ credible intervals.}
    \label{fig:gmc-ysc-estimate}
\end{figure*}

\begin{figure*}
\centering
\subfloat[Residuals of the model]{
\includegraphics[width = 80mm]{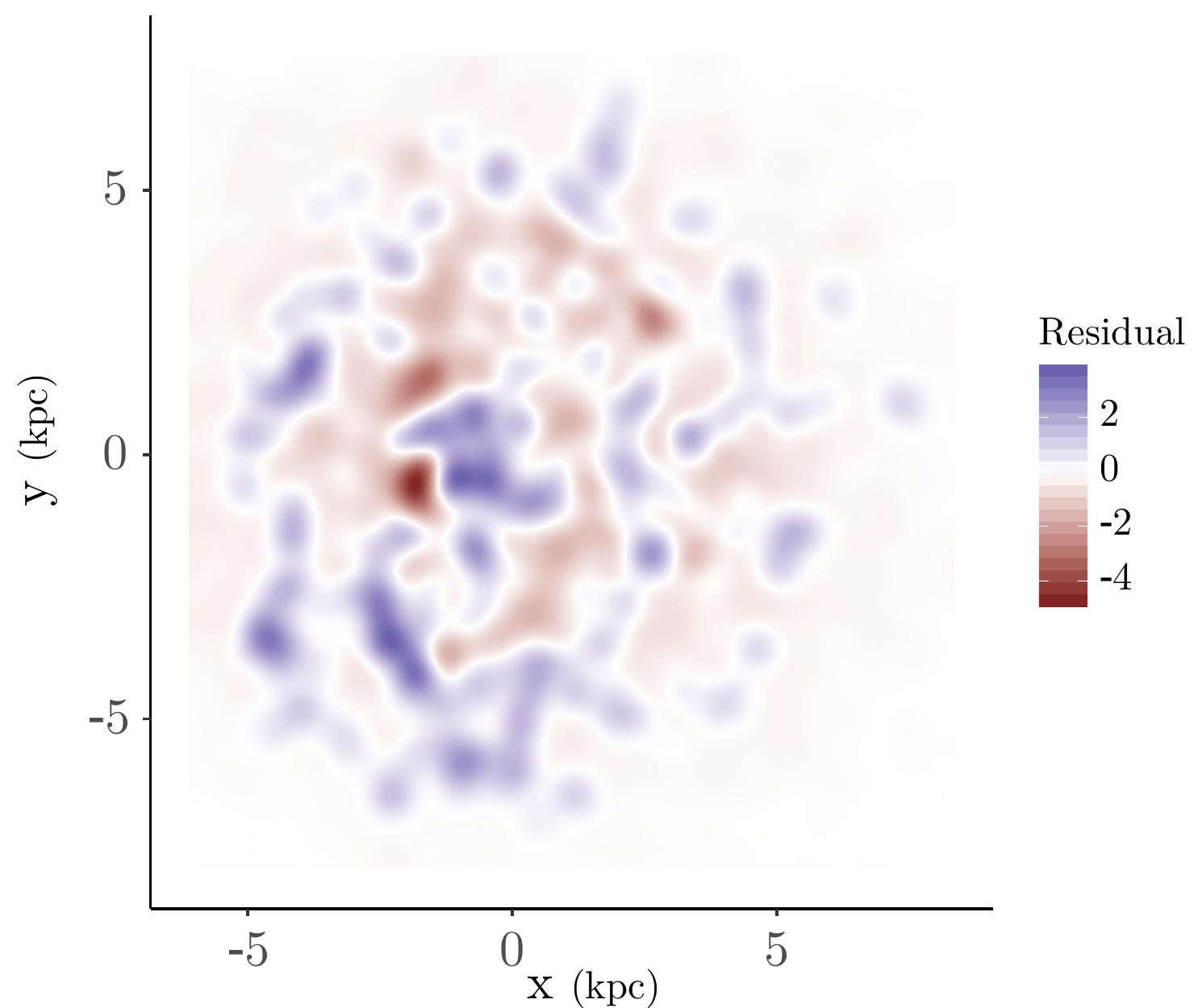}
}
\hspace{10mm}
\subfloat[$95\%$ pointwise credible intervals coverage]{
\includegraphics[width = 80mm]{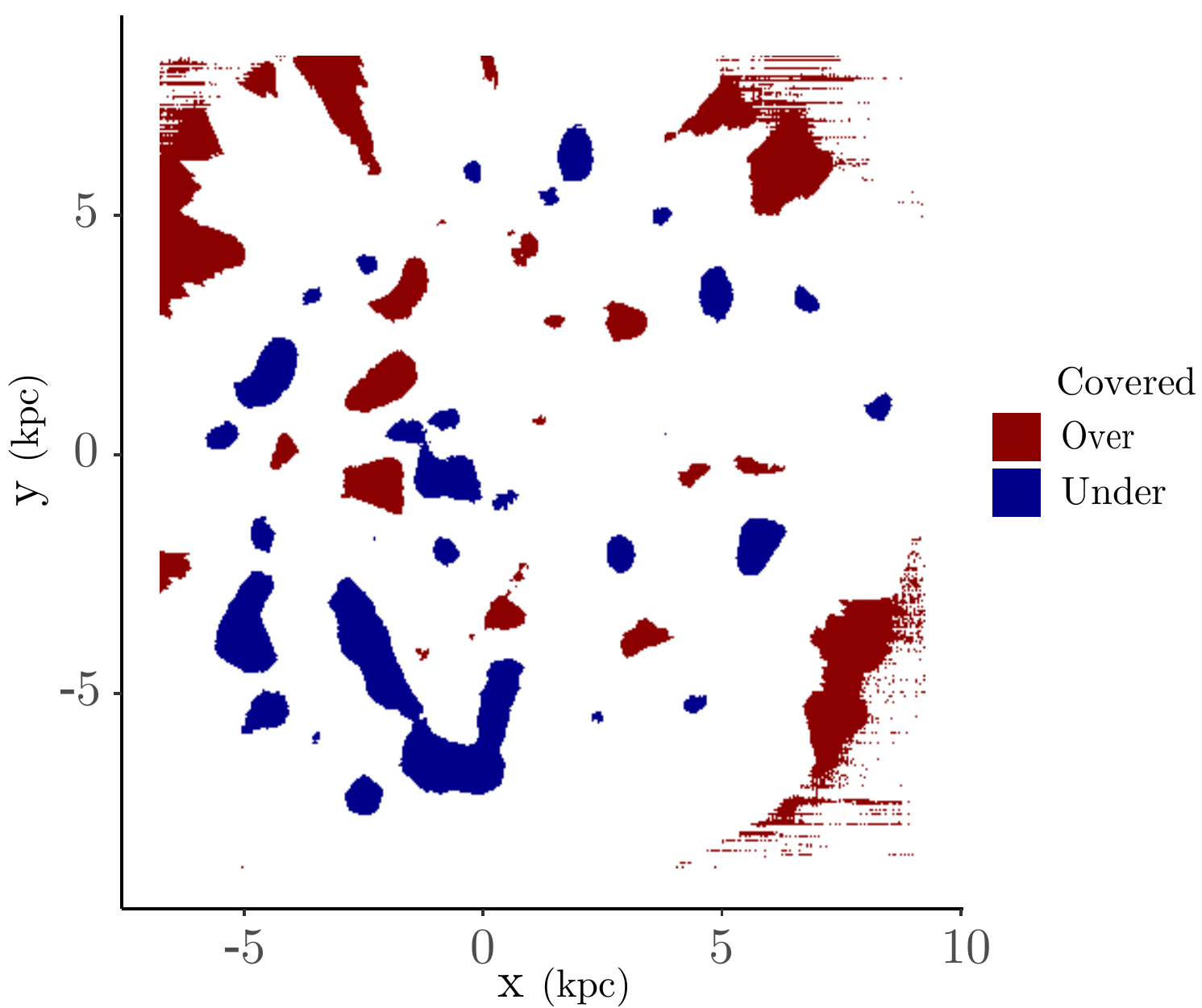}
}
\caption{(a) Residuals obtained from kernel density estimations of the intensity of data and the intensity of 200 posterior simulation. (b) $95\%$ pointwise credible intervals coverage; dark red shows the regions where the $95\%$ credible intervals of residuals are below zero, i.e., the model consistently overestimates the intensity; Dark blue shows the regions where the $95\%$ credible intervals of raw residuals are above zero, i.e., the model consistently underestimates the intensity; White shows the regions where the $95\%$ credible intervals of raw residuals cover zero.}
\label{residuals}
\end{figure*}

From Figure \ref{fig:model_validate}, we can conclude that our procedure can indeed recover the true parameters with reasonably good performance as most 95\% credible intervals cover the true parameters. 
From results obtained on simulated data, we can proceed to conduct data analysis on the real data using the constructed model and the DMH algorithm.

\section{Data Analysis and Model Criticism} \label{sec:data}

\subsection{Prior Setup \& Posterior Results}
To set up the Bayesian framework, ten independent chains are run with prior distribution chosen to be $\mathcal{N}(\boldsymbol{\theta}^0, 100^2\mathbf{I})$, i.e., a multivariate normal distribution with mean vector $\boldsymbol{\theta}^0$ and covariance matrix $100^2\mathbf{I}$. $\boldsymbol{\theta}^0$ are crude estimates obtained from preliminary summary statistics and $\mathbf{I}$ is the identity matrix.  Table \ref{tab:gmc_sc_crude} provides the values chosen for $\boldsymbol{\theta}^0$. The crude estimate of $R_{s,c}^0$, for example, is chosen to be 5~kpc as observed in Figure \ref{hist_gmc_sc_d} and $\sigma_{GS}^0$ is set as 76~pc based on the median nearest neighbour distance from YSCCs to GMCs. The crude estimates for other parameters such as $\rho$ or $\theta_0$ are obtained through trial and error by comparing the simulated data and real data. For parameters such as $\theta_D$, it is difficult to obtain a crude estimate and they are therefore set to zero. The variance of each parameter is set to $100^2$ to reflect the large uncertainty of our crude estimates. 100,000 iterations are carried out for each chain. The parameters whose domain is strictly positive are transformed into log-scales. For simplicity, we set the size parameter of YSCCs $R_S$  to 10~pc which is close to the average size of YSCCs in the data. To boost the convergence speed, an adaptive MCMC scheme \citep{haario_adaptive_2001, roberts_examples_2009, brooks_optimal_2011} is employed, where the proposal distribution of DMH algorithm is set to the following:
\begin{equation}
    q(\boldsymbol{\theta}'|\boldsymbol{\theta}) \sim \mathcal{N}(\boldsymbol{\theta}, 0.1\boldsymbol{\Sigma}_n + 0.01\boldsymbol{D}).
\end{equation}
$\boldsymbol{\Sigma}_n$ is the covariance matrix of the first $n$ values of the chain and $\boldsymbol{D}$ is a user-defined diagonal matrix with small diagonal elements to ensure the total covariance matrix of $q(\cdot|\cdot)$ is invertible. Convergence diagnostic plots of the DMH algorithm are given in Appendix \ref{convergence_daignose}. We also conducted a sensitivity analysis on the effect of changing the crude estimates for the prior distributions and found that the change in the resulting posterior distributions is rather minute. For example, a change in the crude estimate of $\sigma^0_S$ from $100$ pc to $1000$ pc resulted in the posterior mean of $\sigma_S$ only increasing from $80$ pc to $85$ pc.

\begin{figure*}
\centering
\subfloat[Empirical PCF for YSCCs]{
\includegraphics[width = 80mm]{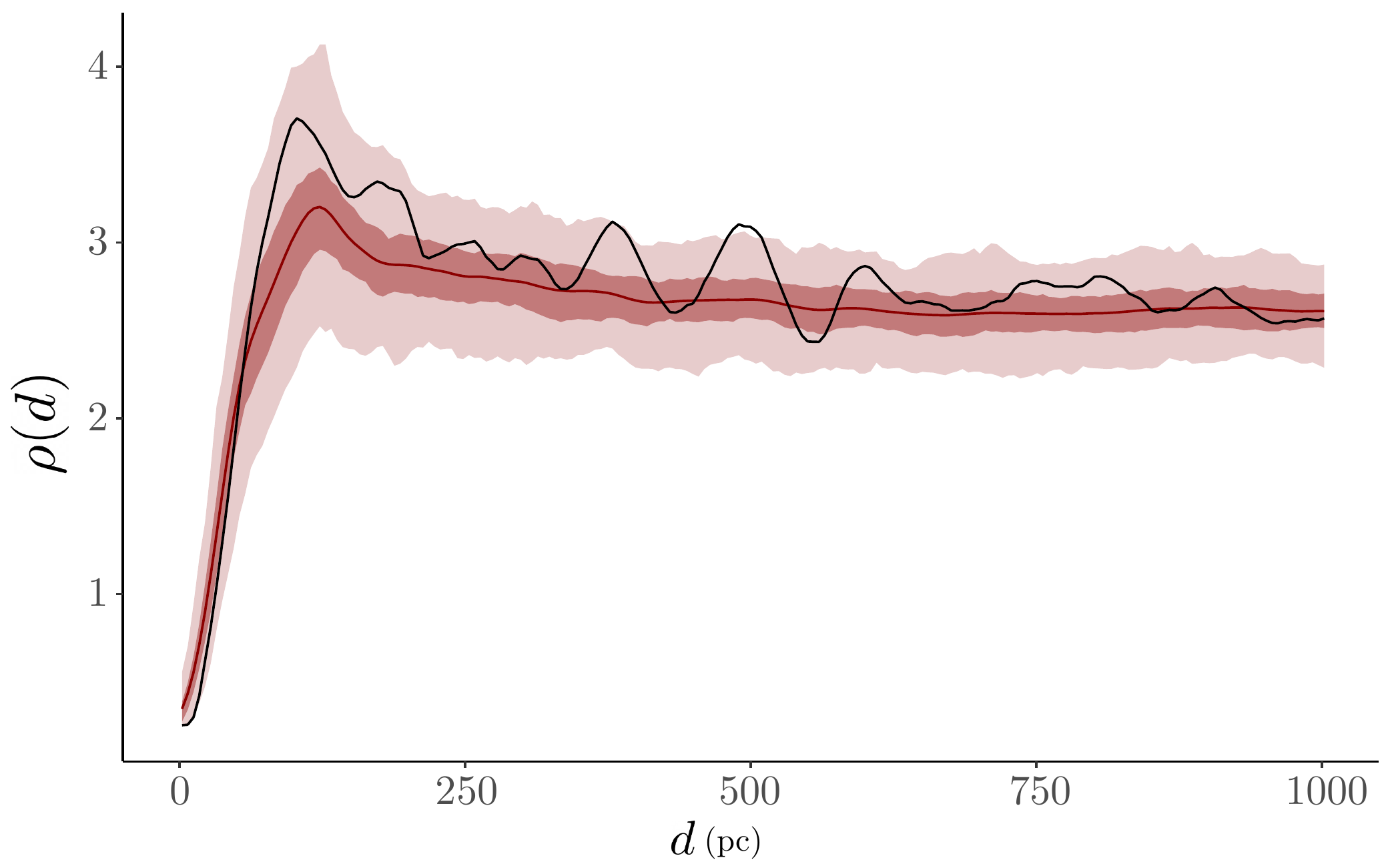}
}
\hspace{10mm}
\subfloat[cNND distributions for YSCCs]{
\includegraphics[width = 80mm]{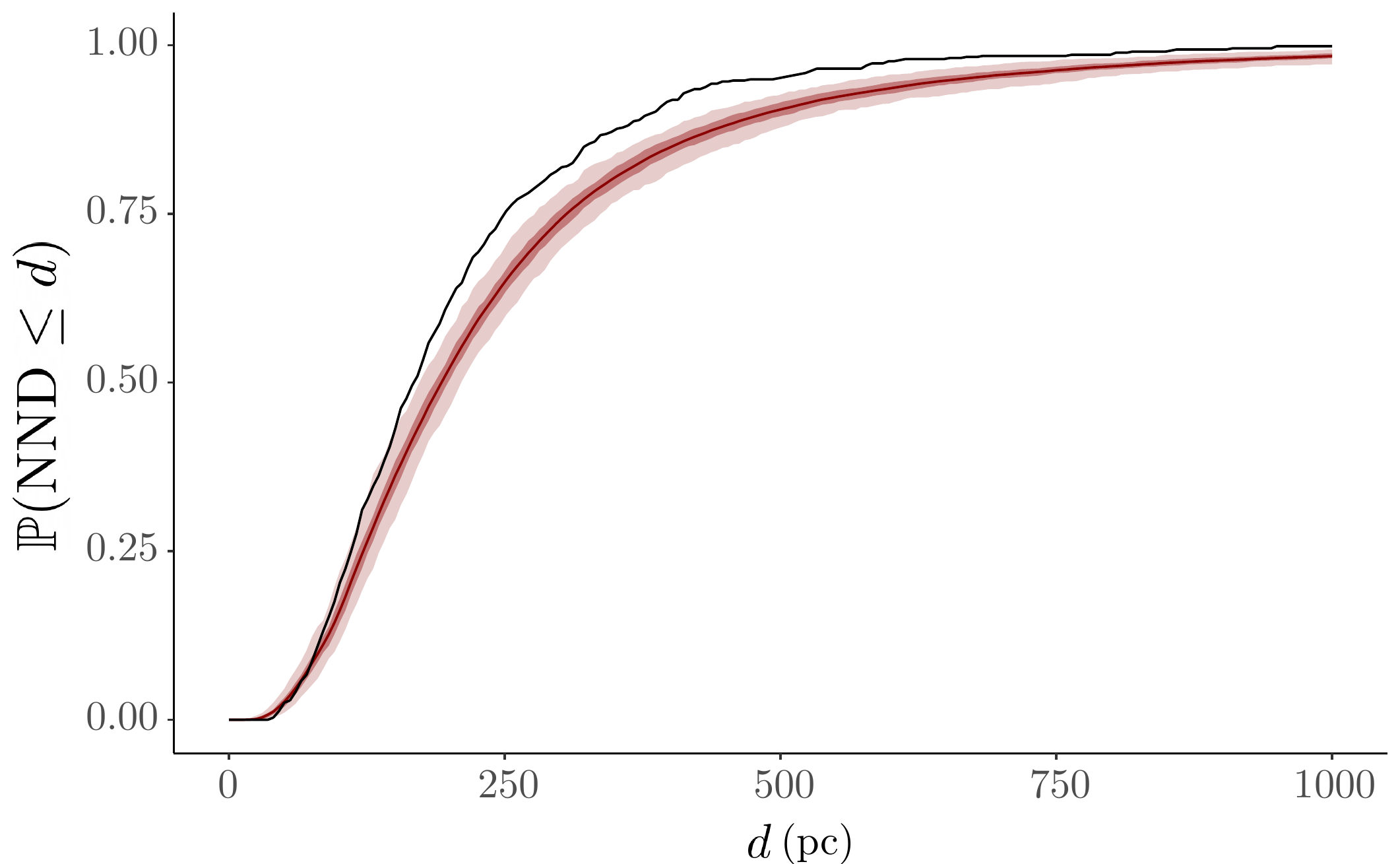}
}
\caption{(a) PCF comparison between data and GMC-SC model: black line is the PCF obtained from data; red line is the estimated mean PCF under the model obtained through 200 posterior simulation; dark red band is the pointwise 50\% credible intervals of the PCF at each $r$ under the model while the light red band is the pointwise 95\% credible intervals of the PCF at each $r$ under the model. (b) cNND distribution function for YSCCs: red line is the estimated mean cNND distribution function from 200 posterior simulations; black line is the cNND distribution function estimated from data; dark red band is the $50\%$ credible band of the cNND distribution function for the model while the light red band is the $95\%$ credible band.}
\label{2ndstats}
\end{figure*}

Figure \ref{fig:gmc-ysc-estimate} provides the corner plot of the posterior distribution as well as summary information for the posterior distribution and estimates. We see that the characteristic scale of the YSCCs in the galactic plane, represented by $R_{s,c}$, is $\sim 4.8$~kpc from the centre of the galaxy. The central intensity, $\rho$, controlling the galaxy-wide first-order potential of the distribution of the YSCCs, is only about 0.7, significantly less than $\theta_0$, the first-order potential contributed by the GMCs which is around 4.5. The characteristic scale, $\sigma_{GS}$, of the correlation between GMCs and YSCCs is $\sim 85$~pc. For the values of $\theta_{D}, \theta_{M}, \theta_{gc}$, we have scaled the properties' data by standardizing with respect to the standard deviation of each property. The effect of galactocentric distance of GMCs $D$, represented by $\theta_{D}$, indicates that if the distance increases by 1 standard scale, the correlation between GMCs and YSCCs increases by $\exp(0.86)$, while 1 standard log-scale increase in the mass of the GMC leads to an increase in correlation strength by about $\exp(0.66)$. The effect from the distance between GMCs and CO filament, however, does not seem to have significant effect on the correlation between GMCs and YSCCs. It is still interesting that $\theta_{gs}$ has a $62\%$ chance of being negative based on the approximate posterior distribution. In terms of the second-order intensity, the characteristic scale $\sigma_S$ is $\sim 79$~pc. Then according to the model, this means that, on average, the interpoint interaction between two YSCCs disappears, i.e., the point pattern becomes Poisson, once their distance is greater than $\sim 105$~pc.Figure \ref{fig:gmc-ysc-estimate} shows that $\sigma_S$ exhibits a bimodal distribution. This is most likely due to the choice of model structure rather than there actually being two potential values for $\sigma_S$. Since the second-order intensity is only first-order differentiable, it is not sufficiently smooth. It also makes no physical sense to have two possible values for a scale parameter. We defer the detailed discussion of the physical implications to section \ref{sec:discussion}.

\begin{figure}
    \centering
    \includegraphics[width = 85mm]{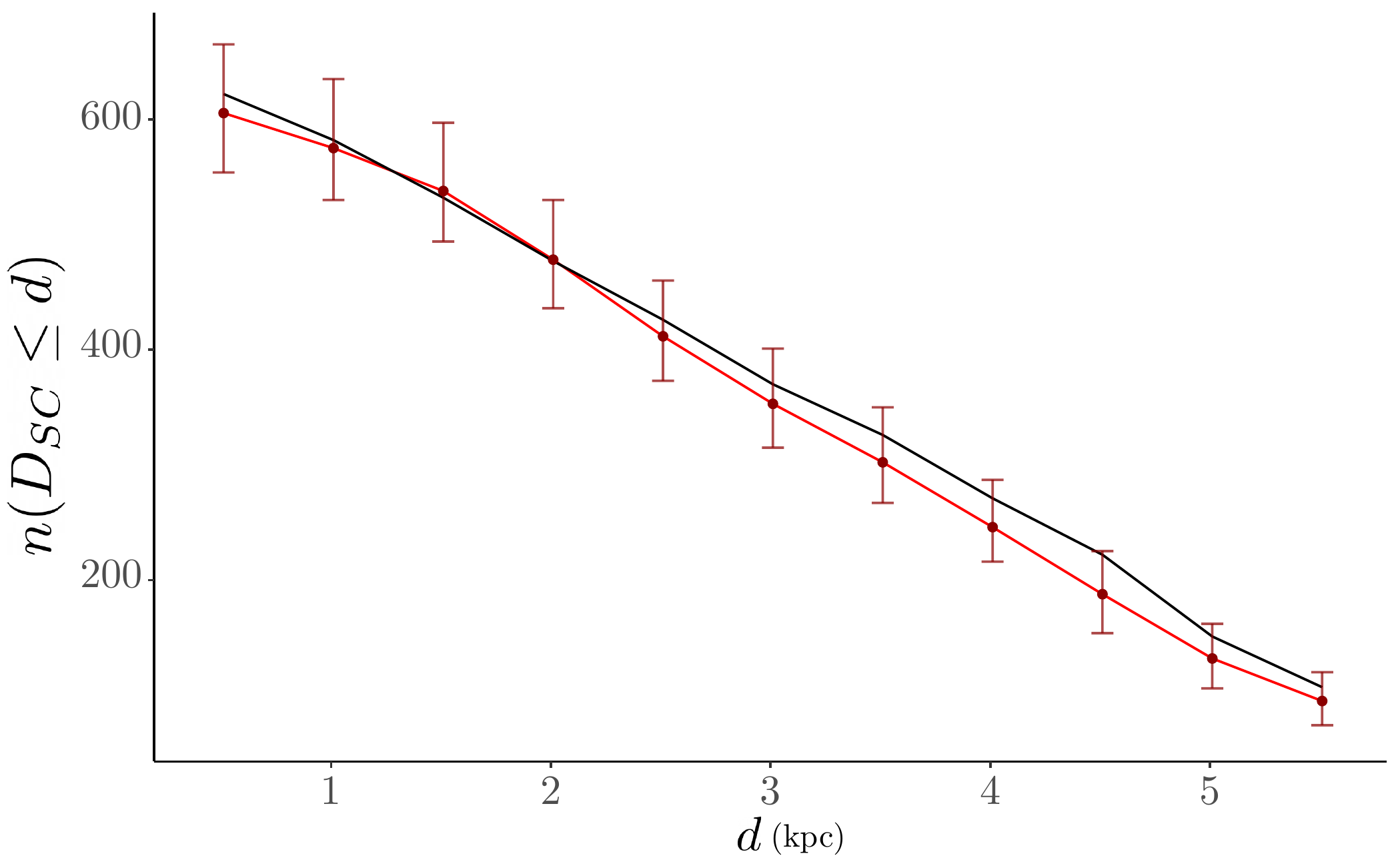}
    \caption{Count of points that are at least distance $d$ away from the galactic centre with $d$ increasing from $0.5$ kpc to $5.5$ kpc; red line and dots are the mean count obtained from simulated data using 200 posterior samples; dark red vertical lines are $95\%$ credible intervals of the count at each distance of $d$ where the count is calculated; black line is the true count at each $d$ where the count is calculated.}
    \label{count_validate}
\end{figure}

\begin{figure}
    \centering
    \includegraphics[width = 85mm]{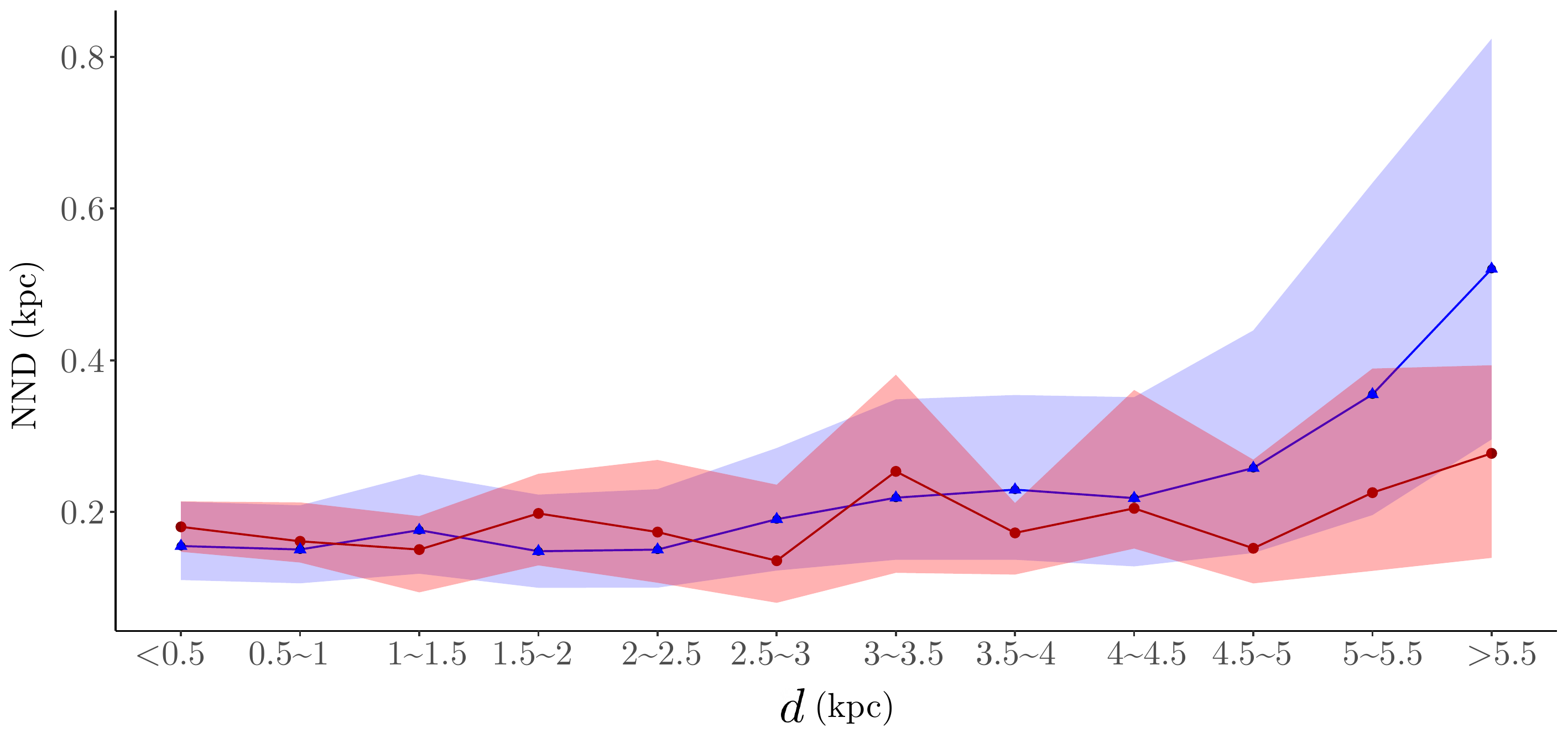}
    \caption[50\% credible intervals of nearest neighbor distances (NND) of YSCCs grouped by distance to the galaxy centre.]{50\% credible intervals of nearest neighbor distances (NND) of YSCCs grouped by distance to the galaxy centre. Red band denotes the central 50\% confidence intervals of NND for each annuli obtained from data; red dots are the median NND from data; blue band denotes the central 50\% credible intervals of NND for each annuli obtained from 200 posterior simulations; blue triangles are the median NND from the 200 posterior simulations.}
    \label{nnd_annulus}
\end{figure}

\begin{figure*}
    \centering
    \includegraphics[width = 100mm]{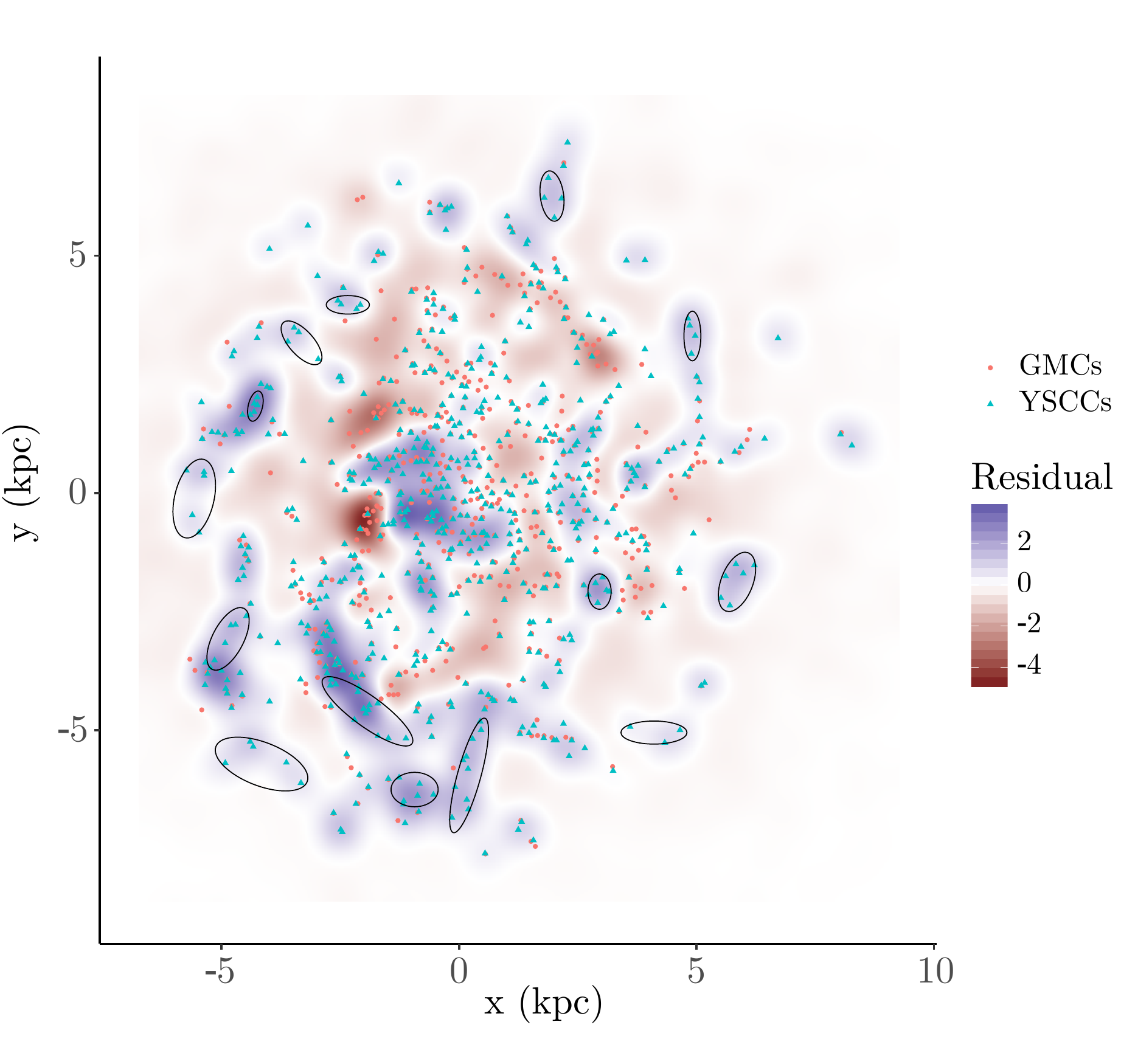}
    \caption{GMCs and YSCCs overlaid on residuals between data and model; Ellipses in the plot show the regions where the intensity is underestimated by the model and  there are no/disproportionately few GMCs in the vicinity of YSCCs.}
    \label{gmc_sc_overlay_residuals}
\end{figure*}

\subsection{Residuals \& Second-Order Structure}
For model criticism, Figure \ref{residuals} shows the intensity residuals obtained by comparing the data and simulation from the fitted model using 200 posterior samples. The intensity residuals are analogous to the residuals obtained from a linear model and are highly useful in diagnosing the model fit. We  do this by employing a kernel smoothing technique to obtain a smoothed residual field with a selected kernel centred at each point. The theoretical background of the method under MLE setting is provided in \cite{baddeley_residual_2005}. Since we are adopting a Bayesian approach, we will implement the method as specified in \cite{leininger_bayesian_2017}.

The continuous map of residuals is computed using a 400$\times$400 grid using the package \textbf{spatstat} \citep{baddeley_spatial_2015} in R \citep{R_2018}. A Gaussian density is used as the smoothing kernel. For bandwidth selection, we found that a bandwidth of approximately 420~pc to be a reasonably good choice through cross-validation. Figure \ref{residuals} (a) shows the mean posterior predictive residuals for the intensity. In general, the residuals are close to 2D white noise across the observation window, indicating a reasonably good fit of the model to data. Figure \ref{residuals} (b) shows the $95\%$ pointwise credible intervals coverage. It shows that the intensity residuals are indeed close to zero in most region of the observation window, confirming a reasonable fit of the model to data. The consistent overestimation in the corner regions of the observation window should be ignored as there are no points in these regions in the data. Therefore, as long as simulation produced points by chance in these regions, it is unlikely that zero will not fall into the $95\%$ credible intervals of residuals for these regions. We see from Figure \ref{residuals} (b) an interesting result in that the intensity in the outer region of the galaxy is consistently underestimated, denoted by the large blue blocks in the plot. This can potentially have multiple explanations and we will need other model diagnostics to pinpoint the possible cause.

To further our diagnostics, Figure \ref{2ndstats} presents the empirical second-order summary statistics of YSCCs using both the empirical PCF and the cumulative nearest-neighbour distance (cNND) distribution. Figure \ref{2ndstats} (a) shows the empirical PCF between the real data (black line) and the mean PCF (red line) from the same 200 posterior simulated data used for Figure \ref{residuals}. Figure \ref{2ndstats} (b) shows the the cNND distributions between data and model. From the plot for empirical PCF, it seems to suggest that the model fits the data reasonably well in terms of the second-order structure of the point pattern. As indicated in the plot, the empirical PCF obtained from data is well within the 95\% credible band obtained from the posterior simulations. However, Figure \ref{2ndstats} (b) shows that in terms of NND distribution, the model demonstrates a strong discrepancy with data. Figure \ref{2ndstats} is in fact an illustration of how being solely dependent on PCF/2PCF may lead to problematic conclusions. This is because PCF/2PCF is in essence a description of the second-order structure through the aspect of correlation between points. It unavoidably possesses a blind spot when it comes to having a complete picture of the second-order structure of a point pattern. NND distribution, on the other hand, is a description of the second-order structure through the aspect of spacing between points, which PCF/2PCF cannot effectively capture. Therefore, a well-rounded second-order analysis through the use of different summary statistics is a much desirable approach. 

From Figure \ref{2ndstats} (b), we see that over short ranges ($r < 100$ pc), the cNND distributions of data and model match reasonably well. However, starting from approximately 150 pc, the point pattern from the data becomes more clustered than the model, peaking at around 250 pc with a difference of 0.1. This means, on average, a YSCC from the data has an excess of 10 percent chance to that of the model of finding another YSCC as its neighbour within 250 pc. This discrepancy of clustering behaviour then declines but remains non-zero all the way to over 600 pc.

It is crucial to point out that neither the empirical PCF or the cNND distribution presented here are obtained after correcting for the inhomogeneity in the point pattern. However, since our goal here is to check how well the model captures the data, it is acceptable to employ them without correcting for inhomogeneity. However, care should be taken regarding the interpretation of discrepancies of the summary statistics between data and model. One detail we need to notice is that since the inferred inhibitive range $R_P \approx 105$ pc while the observed clustering roughly starts at $\sim 150$ pc, the clustering feature observed in Figure \ref{2ndstats} (b) is indeed occurring with respect to a Poisson structure rather than the inhibitive structure at short range. Combining the findings from Figure \ref{residuals} (b), we can conclude that this discrepancy originates from the underestimated blocks in the outer region of the galaxy. However, there are three potential causes for this underestimation due to potential model misspecification: (1) the underestimation of the large-scale inhomogeneity in the outer region; (2) the underestimation of the correlation strength with GMCs; (3) second-order clustering not accounted for by the model. We carry out further analyses in the next section to investigate the likelihood of the three potential causes listed above.

\subsection{Goodness-of-Fit for Large-Scale variation \& GMC-YSCC correlation}
To see the general estimates of the large scale effect, Figure \ref{count_validate} shows a count comparison between the data and the model with respect to the distance from the centre of the galaxy to its outer rim. We do this by counting the number of points within a region  distance $d$ away from the galaxy centre, where $d$ ranges from 0.5 kpc to 5.5 kpc in 0.5 kpc increments. We compare the statistics from the data to what is obtained from simulation of 200 posterior samples. Figure \ref{count_validate} shows that the data and the model are generally in good accordance with each other, meaning that the large scale inhomogeneity is indeed sufficiently accounted for. 

Furthermore, using the same simulated data obtained for Figure \ref{count_validate}, we plotted the comparison of the NND distribution of YSCCs in annuli encompassing the galaxy centre. The result is shown in Figure \ref{nnd_annulus}. The discrepancy between the NND distribution in each annulus is reasonable until the annuli start to reach the outer region of the galaxy, starting from $d > 4.5$ kpc. Furthermore, the median NND distance of YSCCs in the outer region are generally close to 250 pc, which corresponds exactly to the distance at which the peak of discrepancy is reached in the cNND distribution in Figure \ref{2ndstats}. 

The conclusion we can obtain from Figure \ref{count_validate} and Figure \ref{nnd_annulus} is the following: the underestimation of the overall intensity in the outer region is not due to mis-specification of the large-scale intensity variation. Rather it is because that points in the data group together more often than in the model.

To determine whether this discrepancy is due to the underestimation of correlation with GMCs, we present the following figures.
Figure \ref{gmc_sc_overlay_residuals} shows an overlay of GMCs and YSCCs on top of the residuals from Figure \ref{residuals}(a). Figure \ref{gmc_sc_overlay_residuals} shows that in the outer rim, the regions where the intensity is consistently underestimated in fact have no or disproportionately few GMCs in their vicinity. We determine the vicinity by referencing the estimated characteristic scale $\sigma_{GS}$ between GMCs and YSCCs which is only about 85 pc. We also marked the regions with no or few GMCs in their surroundings with ellipses for better visualisation. These ellipses correspond to the regions where intensity is consistently underestimated in Figure \ref{residuals}(b). 

\begin{figure}
\centering
\subfloat[Real $R_{gs}$ vs $R_{ss}$]{
  \includegraphics[width=80mm]{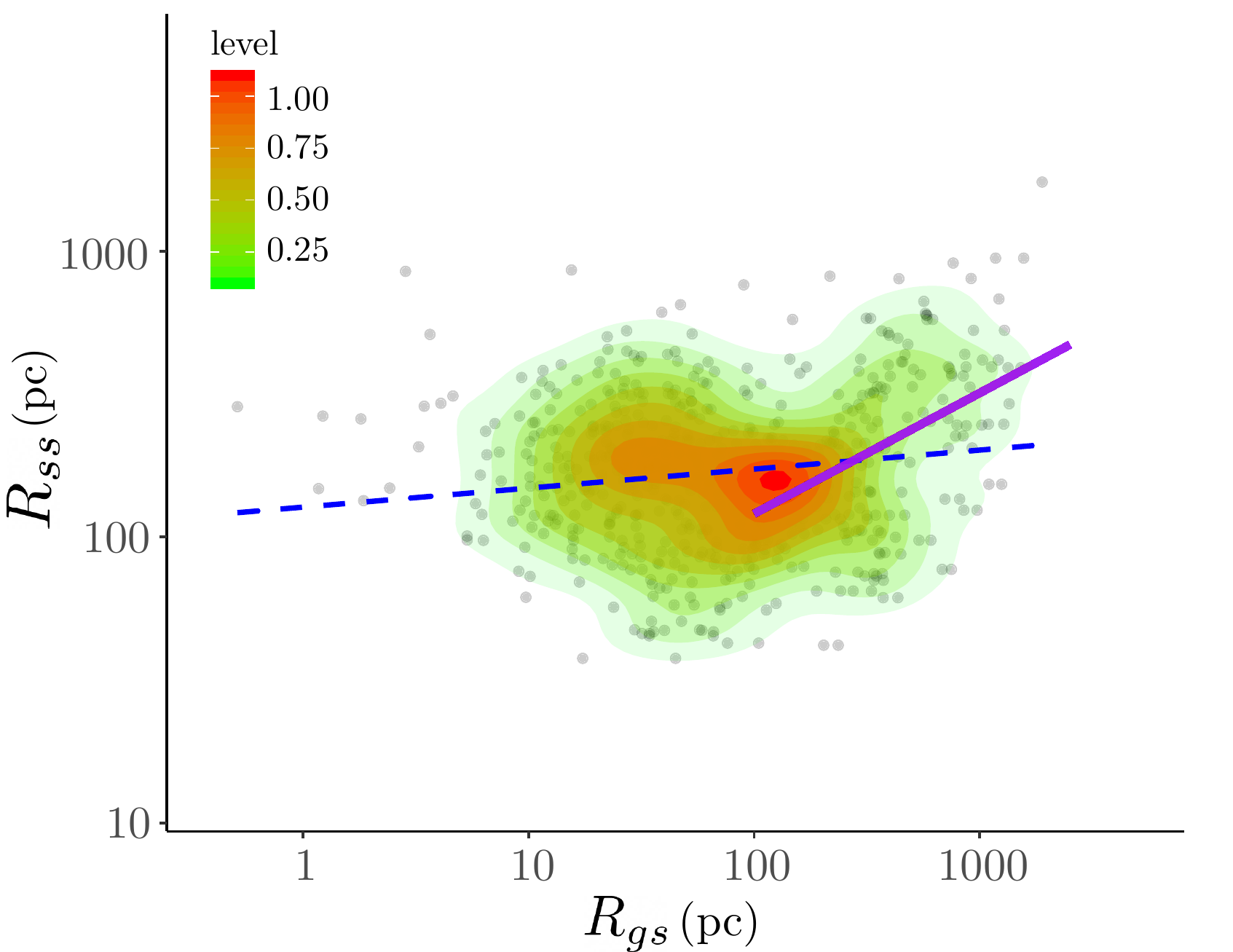}
}
\hspace{0mm}
\subfloat[Posterior simulated $R_{gs}$ vs $R_{ss}$]{
  \includegraphics[width=80mm]{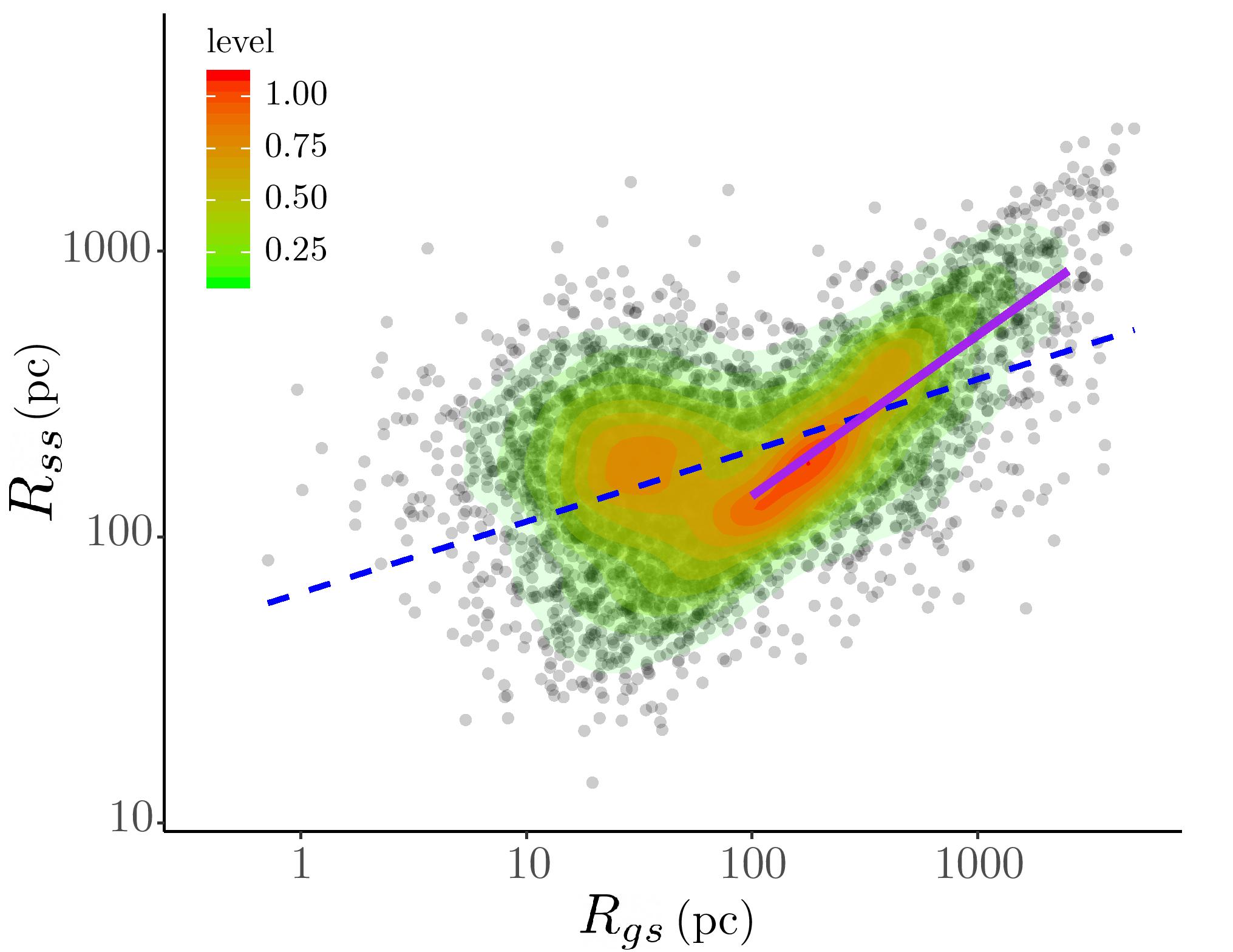}
}
\caption{Density contours of distance from YSCCs to nearest neighbour in GMCs ($R_{gs}$) against the nearest neighbour distance between YSCCs ($R_{ss}$); (a) Plot obtained from real data; (b) Plot obtained from 200 posterior simulations; Dashed blue lines are the fitted least square lines between the two distances; Solid purple lines are the fitted least square lines between the two distances given $R_{gs} > 100$~pc. The plots are in log-log scale.}
\label{density_nn_gs_ss}
\end{figure}

For a more quantitative inspection, we also plot the bivariate density between the distance from a GMC to its nearest neighbour in YSCCs ($R_{gs}$) against the distance from that YSCC to its nearest neighbour in YSCCs ($R_{ss}$). Figure \ref{density_nn_gs_ss} shows that there is a huge discrepancy between the data and the model when $R_{gs} > 100$ pc; however, there is not much discrepancy at $R_{gs} < 100$ pc. The blue-dashed line in the plots are the fitted least-squares line between $R_{gs}$ and $R_{ss}$. For the observed data, the slope of the line is 0.06 while it is 0.25 for the simulated data. The purple lines are fitted least-squares lines given $R_{gs} > 100$ pc. The slope for the real data is about 0.42 and the slope is 0.56 for the simulated data. From this, we can determine that the point pattern in the data is in fact more clustered than the simulated data from the model when the YSCCs considered are far away from the GMCs. Furthermore, given that this discrepancy occurs mainly at range $R_{gs} > 100$ pc and peaks at around 250--300 pc, we can conclude that the underestimation of intensity of YSCCs in the outer region is not caused by the underestimation of their correlation with GMCs. From a physical sense and from the inferred value of $\sigma_{GS}$, the influence of GMCs on YSCCs should not extend to over 250~pc. 
Therefore, combining all the results from previous analysis, we see that there is indeed second-order clustering unaccounted for by the model at 250--600~pc scales and this occurs in the outer region of the galaxy.

\section{Discussion} \label{sec:discussion}

\subsection{First-Order Potential and Correlation Structure}

The parameters governing the first-order potential provide some very interesting insights on the star formation process in M33. The characteristic galactocentric distance, represented by $R_{s,c}$, is $\sim4.8$~kpc. This coincides well with the mean of the prior distribution for $R_{s,c}$ at 5~kpc. The central intensity, $\rho$, controlling the galaxy-wide first-order log-intensity of the distribution of the YSCs is only about 0.68. This means that at the centre of the galaxy, the first-order intensity contributed by the large-scale intensity is approximately $\exp(0.69) = 1.97$~kpc$^{-2}$. This can be explained as approximately 2 YSCCs per kpc$^2$ at the galaxy centre occurring not due to the presence of GMCs, rather the general intensity variation across the galaxy disc. This number will then drop as one moves away from the galaxy centre. In the immediate surroundings of a GMC, the baseline correlation strength parameter $\theta_0$, or the first-order log-intensity contributed by an average GMC is around 4.5. This means that at the same galactocentric distance, the increase in the intensity from a region with no GMC to the centre of an average GMC is a walloping $\exp(4.5) = 90$ times. This confirms that there indeed is a strong correlation between GMCs and YSCCs as suggested by \cite{corbelli_molecules_2017} and it provides rigorous proof that this correlation between GMCs and YSCCs is not simply due to the general overlapping distribution among them across the galaxy disc. This also provides evidence to suggest GMCs are indeed the birthplaces of YSCCs since a correlation strength at this level is highly unlikely to be due to random alignment between GMCs and YSCCs.

However, $\rho = 0.69$ does not equate to saying the overall intensity contributed by the large scale first-order intensity is 2 YSCCs per kpc$^2$ at the galaxy centre. Rather, we do not know the overall intensity as it is also governed by the second-order intensity. However, the increase in the overall intensity from regions with no GMC to the vicinity of an average GMC is indeed 90 times.

The characteristic scale $\sigma_{GS}$ of the correlation between GMCs and YSCs is about 85~pc; this matches well with the median distance of $76$~pc from a GMC to its nearest neighbour in YSCCs. A slightly greater estimated value is largely due to the fact that we considered all possible assignments of a YSCC to a GMC. It is also similar to the general size of cloud-scale star formation complexes \citep[$\lesssim 100$~pc; ][]{chevance_lifecycle_2019}. However, compared to the correlation scale of $17$~pc obtained by \cite{corbelli_molecules_2017}, the difference is rather drastic. The reason for the drastic differences between the two approaches are due to completely different methodologies. In \cite{corbelli_molecules_2017}, the correlation scale parameter is obtained by utilizing the distance from a GMC to its nearest YSCC, while our method uses distances of all GMC-YSCC pairs. It is then clear why our estimate is drastically higher than their estimate since the two estimates carry completely different physical meanings. As to which method is superior, there is no definitive conclusion since two methods are characterizing the relationship of GMCs and YSCCs through different lenses: \cite{corbelli_molecules_2017} used nearest neighbor distance to characterize the spatial relationship through the notion of spacing while our method does it by describing the spatial correlation between two point patterns. Nevertheless, a characteristic scale of 85~pc still shows a strong positive correlation between GMCs and YSCCs. Furthermore, it also means that the correlation strength between GMCs and YSCCs diminishes drastically as the separation distance increases.

For the slope parameters governing the effect of GMC properties on the correlation strength with YSCCs, we found that $\theta_D = 0.86$, $\theta_M = 0.66$, and $\theta_{gc} = -0.05$.
The value of $\theta_D$ shows that the correlation strength increases by $\exp(0.86) = 2.3$ if the galactocentric distance of GMCs increases by 1 standard scale, which is about 1.55 kpc. This is consistent with the preliminary analysis on the cross-type PCF between GMCs and YSCCs obtained in Figure~\ref{cpcf}. To better compare our results to those of \cite{corbelli_molecules_2017}, we follow the procedure described in that work and analyse the ratio between the ``positional correlation function'' of GMCs and YSCCs in three radial zones. We found that the maximum increase in the ratio is around 3 when moving from zone 1 ($D < 1.5$ kpc) to zone 2 ($1.5$ kpc $ \leq D < 4$ kpc) and about 2 from zone 2 to zone 3 ($D \geq 4$ kpc). This is generally in line with what we have obtained, although differences in estimates diverge as the galactocentric distance increases. Again, this is likely due to the completely different approach in modelling since for simplicity, we considered the effect of galactocentric distance on the correlation strength as linear across the galaxy disc, which could be unrealistic. We will consider other forms of non-linear relationships in future work.

Interpreting the physical meaning of $\theta_D$ is complicated since many properties of spiral galaxies change with galactocentric distance and could potentially affect the strength of the correlation between GMCs and YSCCs. However, the most probable cause is the change in tidal shear with respect to the galactocentric distance.
Tidal shear due to differential rotation will separate YSCCs and their birth GMCs more quickly in the inner parts of galaxies. Tidal shear can also unbind GMCs, making it more difficult for them to form clusters in the first place, or destroy clusters after they are formed.
The interstellar radiation field and cosmic ray density also change with galactocentric distance, but a physical mechanism that could cause them to affect GMC-YSC correlations is not as apparent. 
The value of $\theta_D$ can also be affected by the association of GMCs and YSCCs  with the galaxy's spiral arms. This might be a potential lurking variable that could influence the actual correlation between GMCs and YSCCs as noted by \cite{corbelli_molecules_2017}.  We do not pursue the modelling with spiral arm structure since that would drive up model complexity and the model considered here already has eight parameters.

The strong positive effect of galactocentric distance on the correlation strength between GMCs and YSCCs leads us to make an important observation. As we have already seen in Figure \ref{gmc_sc_overlay_residuals}, the outer region of the galaxy disc has a number of groups of YSCCs. Although we have pointed out that these groups do not have GMCs in their immediate surroundings ($< 100$~pc), a partial contribution to the high value of $\theta_D$ could come from the fact that these YSCC groups all appear to be within 200--500~pc of GMCs. We argue that this should not be caused by the crowding between GMCs and YSCCs in the spiral arms since (a) the scale of 200--500 pc is still relatively local for spiral arms to have any significant effect on the density variations of both GMCs and YSCCs; (b) YSCCs  need to have a birthplace and they cannot show up out of nowhere simply because of the presence of spiral arms. The point of this observation is that these YSCC groups not having GMCs in their surroundings at a distance on the order of $\sigma_{GS}$ may have important physical implications for the formation and evolution of YSCCs, further discussed in section \ref{subsec: sec-order}.

The value of $\theta_M$ shows that the mass of GMCs also has a strong positive effect on the correlation strength between GMCs and YSCCs. Similar to the effect of the galactocentric distance, 1 standard scale ($2.1 \times\log_{10}(M_\odot)$) increase in the mass of a GMC can lead to a $\exp(0.66) = 1.9$ times increase in the correlation strength. This also corresponds to the finding in \cite{corbelli_molecules_2017} where they noted that $69\%$ of the high-mass GMCs ($> 2\times10^5 M_\odot$) have a YSCC within $50$~pc while only $44\%$ of low-mass GMCs have an associated YSCC. 

The distance from GMC to the CO filament structure does not seem to have any significant effect on the correlation strength between GMCs and YSCCs. However, as noted in section \ref{sec:data}, the approximate posterior distribution of $\theta_M$ shows that 60\% of the posterior samples are below 0. This, together with the estimated posterior mean at $-0.05$, shows that as GMCs break away from the CO filament, their correlation with the YSCCs tends to slightly decrease. This may indicate that the star formation activity is more fervent while GMCs are still part of the CO filament, although the effect is small.

\subsection{Second-Order Potential} \label{subsec: sec-order}

Based on the second-order potential and the results from model criticism, we confirm that there indeed exists inhibitive behaviour between YSCCs at short distances, as indicated by the matching of the NND distribution at short distances in Figure~\ref{2ndstats}(b). The most important results we found are on the YSCC clustering behaviour in the outer region of the galaxy disc. As mentioned before, these groups of YSCCs  are not associated with any GMCs, but they are still generally within 200--500 pc from GMCs. Several potential explanations that may shed light on the evolution of GMCs and YSCCs and explain the grouping behavior are:
\begin{itemize}
    \item There are undetected GMCs in the outer region of the galaxy
    \item YSCCs in the outer regions destroyed their natal GMCs
    \item YSCCs moved away from their natal GMCs
\end{itemize}

First, the grouping behavior of YSCCs in the outer region of the galaxy can serve as evidence for the hypothesis proposed by \citet{corbelli_molecules_2017}. In their conclusion, they attributed the non-negligible disparity in the numbers of GMCs and YSCCs in the outer region to the presence of GMCs that are below detection limits, with some of the excess YSCCs born from these undetected GMCs. The detection of the grouping behaviour of YSCCs in the outer region in our analysis can support this hypothesis. If we assume similar levels of correlation between the undetected GMCs and YSCCs and some of these YSCCs are still associated with undetected GMCs, then these GMCs will strongly affect the position of the ``unclaimed'' YSCCs and these YSCCs will likely group around the undetected GMCs. However, these GMCs are not present in the data due to the detection limit. Therefore, the model cannot account for their effect on the YSCCs, which is reflected by the grouping behaviour demonstrated in our analysis. Furthermore, the results from Figure \ref{2ndstats} also seem to point in the direction of the undetected GMCs hypothesis. The recent study by \citet{chevance_lifecycle_2019} analysed the cloud-scale star formation complexes (including GMCs and associated SCs) in nine spiral galaxies. They found that the general mean separation distance between individual star formation complexes is roughly $\sim 100-300$~pc. This corresponds to a similar spatial scale to where the peak of discrepancy occurs between NND distributions of the data and our model as shown in Figure \ref{2ndstats}. If these YSCCs are indeed associated with undetected GMCs that are separated by 250~pc on average, then this explains the discrepancy in Figure~\ref{2ndstats}.

To assess the plausibility of the hypothesis of undetected GMCs, we turn to the original paper of \citet{druard_iram_2014} where the GMC observations are reported. The noise map  presented in Figure 6 of \citet{druard_iram_2014} shows that the  noise variation across the galaxy disc is almost negligible. If we compare the region with the highest noise level with the region with underestimated intensity in Figure \ref{residuals}, the high noise region does not have significant overlap with the blue blocks in Figure \ref{residuals}. Furthermore, the high noise region in fact has detected GMCs. If we assume that the CO intensity from GMCs is on a similar level in the outer region, the above comparison does not seem to support the hypothesis of undetected GMCs.

This conclusion is consistent with the analysis of \citet{gratier_2017}, who used the \citet{druard_iram_2014} observations together with measurements of dust continuum and H{\sc i} emission to estimate M33's spatially-resolved gas-to-dust ratio, ``X-factor'' between $\text{H}_2$ and CO, and projected density of CO-dark gas. 
Numerous studies have shown that M33 has a radial metallicity gradient \citep[e.g.]{cioni_metallicity_2009,magrini_2010}: such a gradient could affect 
$X_{\rm CO}$ and thus the detectability of GMCs in the outskirts of M33. 
However, \citet{gratier_2017} found no evidence for radial variation in $X_{\rm CO}$ or for significant CO-dark gas in the outskirts of M33.
To conclusively test the hypothesis of undetected GMCs, targeted high sensitivity CO observations  in the outskirts of M33 are needed. The residual field in Figure \ref{gmc_sc_overlay_residuals} in fact gives a map which can narrow down the region for the pointed observations: they can simply be made at the regions with the most underestimation in the intensity of YSCCs. This is another demonstration of the power of GPP modelling.

As demonstrated in the previous arguments, the hypothesis of undetected GMCs does not seem to hold. In the case that the targeted observation for undetected GMCs turn out to be unsuccessful, other explanations are needed to explain the grouping behavior of YSCCs. We  propose two additional hypotheses alternative to that of undetected GMCs.

Firstly, the grouping behavior of YSCCs can be caused by them destroying their natal clouds. \citet{corbelli_molecules_2017} concluded that GMCs in M33 tend to have a very short lifetime, around 14.2~Myr. \citet{chevance_lifecycle_2019} also estimated the lifetime of GMCs in nine nearby galaxies and found that they average  $\sim10-30$~Myr. They found that, in general, GMCs in these galaxies spend most of their lifetime ($\sim$75-90\%) dormant but quickly disperse in $\sim1-5$ Myr once the stars are formed, likely due to stellar winds. The study of NGC~300 by \citet{kruijssen_fast_2019}  found evidence of a rapid evolutionary cycle among GMCs and star formation, with GMC destruction in less than 1.5 Myr by efficient stellar feedback. 

A simple deduction can be made that if GMCs are of low mass, their destruction should be even more rapid. \citet{corbelli_rise_2019} analyzed the variation in GMC mass with  galactocentric distance and found that the mass of GMCs drops as galactocentric distance increases. They concluded that the presence of high mass GMCs in the inner disc of M33 ($D < 3.9 $~kpc) is likely due to the supersonic rotation of the disc in the inner region where the gas is collected by the spiral arms and forms more massive clouds. However, this is not the case beyond the co-rotation region ($D > 4.7$~kpc) where the much slower rotation results in low mass GMCs. 
The co-rotation distance of 4.7~kpc corresponds to our observation of grouping of YSCCs beyond 4.5~kpc, and if  GMCs in the outer region belong to the low mass class ($\lesssim 10^5 M_\odot$), then a possible explanation for the absence of GMCs might be the formation of YSCCs and their efficient stellar feedback leading to the destruction of their low mass natal clouds. 

The rapid breakout of SCs from their natal clouds compared to the SCs' dispersion speed also adds to the evidence for natal cloud destruction hypothesis. \citet{hollyhead_studying_2015} showed that young massive clusters in M83 generally break out of their natal clouds around 4 Myr. \citet{corbelli_molecules_2017} also analyzed the association between GMCs and another catalog of optically visible SCs by \citet{fan_star_2014} in M33. Those clusters had a wider range of age estimates (from 5 Myr to 10 Gyr) than those considered here. Although the correlations found between these SCs and GMCs were much weaker than the ones found in this study, the correlations are still stronger than those of a Poisson process. This means that the time scale for SCs to disperse into a Poisson-like structure is much longer than the cloud life-time as suggested in previous studies. This indicates that the grouping behaviour of YSCCs in the outer region is potentially a result of YSCCs destroying their natal clouds before they have had time to disperse and appear Poisson-like. To test this hypothesis, we would need age measurements of the YSCCs. Age estimates are only available for 402 out of the 630 YSCCs with a mean estimate at $\sim$5 Myr. The results for GMC dispersal time ($1 \sim 5$ Myr) after star formation from previous studies \citep{chevance_lifecycle_2019, kruijssen_fast_2019} imply that many GMCs might have just been destroyed by the newly formed SCs through stellar winds. This is even more probable if the destruction of low mass GMCs is more rapid than $\sim 1-5$~Myr. However, the age estimates of the YSCCs in M33 are rather imprecise and should not be used to draw quantitative conclusions. 

Another potential process involved in the appearance of the clustering might be that multiple YSCCs are in fact generated by the same GMC. As these YSCCs break out and lose their association with their original GMCs, they might have similar velocity due to their common birthplace. Since they are all in the early stage of their evolution, they tend to move in the same direction before starting to disperse.  Analysis  by \citet{grasha_spatial_2019} suggests that on average SCs in M51 that are not associated with any GMCs are much older ($\sim 50$ Myr) compared to those that are associated with a GMC ($\sim 4$ Myr). Assuming the star formation process is generally universal, this observation can be indirect evidence to support the hypothesis that the YSCCs in the outskirts are moving away from their natal clouds. This hypothesis also tends to explain the fact that most GMCs in the outer region of M33 tend to have low mass as well as the number disparity between GMCs and YSCCs in the outer region. Low mass and number disparity with YSCCs are potential indication that GMCs may have produced all of their YSCCs and are almost at the end of their life-cycle. However, to test this hypothesis, we would need more accurate estimates of the age of YSCCs to analyze the correlation between GMCs and YSCCs as a function of the age of YSCCs. If the association weakens with increasing age, this would serve as evidence in support of the hypothesis. We cannot carry out a meaningful test for this hypothesis with available data due to the low numerical resolution of the available age estimates: of the 402 YSCCs with age estimates, 255 of them have the same estimate ($\log{\rm (age [yr])}=6.7$) and 51 have an estimate of $\log{\rm (age)}=6.8$. Additionally, among these regions, the region with the largest number of YSCCs is only 11 which is relatively small. Moreover, there are still many YSCCs in these elliptical regions with no age estimates: in 6 out of 14 group, around half of the YSCCs are without an age estimate.

In conclusion, the formation of SCs may be a combination of the processes mentioned above and further detailed study needs to be done to paint a clear picture. Nevertheless, the results we have obtained clearly showcase the power of GPP model in its effectiveness and sensitivity to numerically identify detailed structure of highly inhomogeneous point patterns. The identification of groups of YSCCs in the outer region would not be possible using the previous exploratory statistical tools of 2PCF/PCF and its variants, and has led to potential new hypotheses on the evolution of stellar populations.

\section{Summary and Future Work} \label{sec:summary}

In this study, we found that in M33, there are:

\begin{itemize}
    \item a strong positive correlation between GMCs and YSCCs
    \item a strong positive influence of GMC galactocentric distance and mass on the correlation strength between GMCs and YSCCs
    \item clustering patterns of YSCCs in the outer region of the galaxy that are not due to large scale variation or the presence of GMCs. 
\end{itemize}

We introduced the Gibbs point process modelling framework to investigate the spatial distribution of young stellar cluster candidates and giant molecular clouds in M33. We have shown that this is a powerful statistical modelling technique that provides rigorous and integrated streamlined data analysis with the ability to answer multiple interesting questions simultaneously, compared to previous studies where methods employed are limited, fragmented, and prone to information loss. We confirmed the remarkably strong spatial correlation between GMCs and YSCCs and the model rigourously demonstrated that  the correlation is not due to the large scale overlapping distributions of GMCs and YSCCs across the galaxy disc. Furthermore, we found that the galactocentric distances and masses of GMCs impose a strong positive effect on the correlation strength between GMCs and YSCCs. We also showed the sensitivity of the GPP models in the numerical measurement of point pattern behaviour by identifying clustering patterns among YSCCs in the outer region of the galaxy disc. This provided new evidence to support existing scenarios and also shed new light on other possible scenarios for the star formation process. This information would not be available if a traditional method such as the two-point correlation function were employed. 

We see several potential directions for future work. First, it would be interesting to model a second-order non-stationary process for SCs since it may very well be the case that the second-order stationary assumption for SCs is not true. We can do this by looking for correlations between the second-order structure and the properties of GMCs and SCs. Secondly, our approach can serve as a sensitive validation tool for large hydrodynamic simulation of galaxies. Parameters in the model can effectively determine the validity of simulations in comparison to observed galaxies. This would require  obtaining model fits for a wide range of galaxies with different morphology and physical structure and obtain a baseline distribution for each parameter, since comparing individual real galaxies to individual simulated galaxies provides no meaningful conclusion. However, obtaining baseline distributions will be difficult due to the small sample of galaxies with high quality observation of GMCs.  High quality observations of GMCs for more galaxies are required. Lastly, we can also consider fitting the model to other spiral galaxies and investigate how changing properties of different galaxies affect the model parameters. This can potentially lead to new physical insight on the star formation process in different physical environments.

\section*{acknowledgments}
P. B. acknowledges support from NSERC Discovery Grants.
We thank E. Koch and E. Rosolowsky for  permission to use their data and A. I. McLeod, C. Matzner, M. Houde, S. Bonner and C. de Souza for helpful discussions. 
The computation in this work is supported by Shared Hierarchical Academic Research Computing Network (SHARCNET) and we thank  B. Ge for valuable guidance in SHARCNET computational support. 

\section*{Code and data availability}

The code used in this research will be made available by the corresponding authors upon request. For the CO filament data, availability should be requested from E. Koch and E. Rosolowsky at the University of Alberta. The GMCs and YSCCs data are publicly available; please refer to \cite{corbelli_molecules_2017} for details.



\bibliographystyle{mnras}
\bibliography{references} 

\begin{thebibliography}{}
\makeatletter
\relax
\def\mn@urlcharsother{\let\do\@makeother \do\$\do\&\do\#\do\^\do\_\do\%\do\~}
\def\mn@doi{\begingroup\mn@urlcharsother \@ifnextchar [ {\mn@doi@}
  {\mn@doi@[]}}
\def\mn@doi@[#1]#2{\def\@tempa{#1}\ifx\@tempa\@empty \href
  {http://dx.doi.org/#2} {doi:#2}\else \href {http://dx.doi.org/#2} {#1}\fi
  \endgroup}
\def\mn@eprint#1#2{\mn@eprint@#1:#2::\@nil}
\def\mn@eprint@arXiv#1{\href {http://arxiv.org/abs/#1} {{\tt arXiv:#1}}}
\def\mn@eprint@dblp#1{\href {http://dblp.uni-trier.de/rec/bibtex/#1.xml}
  {dblp:#1}}
\def\mn@eprint@#1:#2:#3:#4\@nil{\def\@tempa {#1}\def\@tempb {#2}\def\@tempc
  {#3}\ifx \@tempc \@empty \let \@tempc \@tempb \let \@tempb \@tempa \fi \ifx
  \@tempb \@empty \def\@tempb {arXiv}\fi \@ifundefined
  {mn@eprint@\@tempb}{\@tempb:\@tempc}{\expandafter \expandafter \csname
  mn@eprint@\@tempb\endcsname \expandafter{\@tempc}}}

\bibitem[\protect\citeauthoryear{Baddeley}{Baddeley}{2007}]{baddeley_spatial_2007}
Baddeley A.,  2007, in Baddeley A.,  Bárány I.,  Schneider R.,   Weil W.,
  eds, Lecture {Notes} in {Mathematics}, Stochastic {Geometry}: {Lectures}
  given at the {C}.{I}.{M}.{E}. {Summer} {School} held in {Martina} {Franca},
  {Italy}, {September} 13–18, 2004.
Springer, Berlin, Heidelberg, pp 1--75, \mn@doi{10.1007/978-3-540-38175-4_1}

\bibitem[\protect\citeauthoryear{Baddeley \& Turner}{Baddeley \&
  Turner}{2000}]{baddeley_practical_2000}
Baddeley A.,  Turner R.,  2000, \mn@doi [Aus. NZ J. Stat]
  {10.1111/1467-842X.00128}, 42, 283

\bibitem[\protect\citeauthoryear{Baddeley, Turner, Møller  \&
  Hazelton}{Baddeley et~al.}{2005}]{baddeley_residual_2005}
Baddeley A.,  Turner R.,  Møller J.,   Hazelton M.,  2005, \mn@doi [J. Royal.
  Stat. Soc. B] {10.1111/j.1467-9868.2005.00519.x}, 67, 617

\bibitem[\protect\citeauthoryear{Baddeley, Rubak  \& Turner}{Baddeley
  et~al.}{2015}]{baddeley_spatial_2015}
Baddeley A.,  Rubak E.,   Turner R.,  2015, Spatial {Point} {Patterns}:
  {Methodology} and {Applications} with {R}.
Chapman \& {Hall}/{CRC} {Interdisciplinary} {Statistics}, CRC Press

\bibitem[\protect\citeauthoryear{Bonanos et~al.,}{Bonanos
  et~al.}{2006}]{bonanos_first_2006}
Bonanos A.~Z.,  et~al., 2006, \mn@doi [\apss] {10.1007/s10509-006-9112-1}, 304,
  207

\bibitem[\protect\citeauthoryear{Carlberg \& Pudritz}{Carlberg \&
  Pudritz}{1990}]{carlberg_magnetic_1990}
Carlberg R.~G.,  Pudritz R.~E.,  1990, \mnras, 247, 353

\bibitem[\protect\citeauthoryear{Chevance et~al.,}{Chevance
  et~al.}{2019}]{chevance_lifecycle_2019}
Chevance M.,  et~al., 2019, \mn@doi [\mnras] {10.1093/mnras/stz3525}

\bibitem[\protect\citeauthoryear{Cioni}{Cioni}{2009}]{cioni_metallicity_2009}
Cioni M.-R.~L.,  2009, \mn@doi [\aap] {10.1051/0004-6361/200912138}, 506, 1137

\bibitem[\protect\citeauthoryear{Corbelli et~al.,}{Corbelli
  et~al.}{2017}]{corbelli_molecules_2017}
Corbelli E.,  et~al., 2017, \mn@doi [\aap] {10.1051/0004-6361/201630034}, 601,
  A146

\bibitem[\protect\citeauthoryear{Corbelli, Braine  \& Giovanardi}{Corbelli
  et~al.}{2019}]{corbelli_rise_2019}
Corbelli E.,  Braine J.,   Giovanardi C.,  2019, \mn@doi [\aap]
  {10.1051/0004-6361/201834437}, 622, A171

\bibitem[\protect\citeauthoryear{Cressie \& Wikle}{Cressie \&
  Wikle}{2011}]{cressie_statistics_2011}
Cressie N.,  Wikle C.~K.,  2011, Statistics for {Spatio}-{Temporal} {Data}.
Wiley, Hoboken, N.J

\bibitem[\protect\citeauthoryear{Daley \& Vere-Jones}{Daley \&
  Vere-Jones}{2003}]{daley_introduction_2003}
Daley D.~J.,  Vere-Jones D.,  2003, An {Introduction} to the {Theory} of
  {Point} {Processes}: {Volume} {I}: {Elementary} {Theory} and {Methods}, 2
  edn.
Probability and {Its} {Applications}, {An} {Introduction} to the {Theory} of
  {Point} {Processes}, Springer-Verlag, New York, \mn@doi{10.1007/b97277}

\bibitem[\protect\citeauthoryear{Daley \& Vere-Jones}{Daley \&
  Vere-Jones}{2008}]{daley_introduction_2008}
Daley D.~J.,  Vere-Jones D.,  2008, An {Introduction} to the {Theory} of
  {Point} {Processes}: {Volume} {II}: {General} {Theory} and {Structure}, 2
  edn.
Probability and {Its} {Applications}, {An} {Introduction} to the {Theory} of
  {Point} {Processes}, Springer-Verlag, New York,
  \mn@doi{10.1007/978-0-387-49835-5}

\bibitem[\protect\citeauthoryear{Davis, Miller  \& White}{Davis
  et~al.}{1997}]{davis_galaxy-weighted_1997}
Davis M.,  Miller A.,   White S. D.~M.,  1997, \mn@doi [\apj] {10.1086/304870},
  490, 63

\bibitem[\protect\citeauthoryear{Dejonghe}{Dejonghe}{1987}]{dejonghe_completely_1987}
Dejonghe H.,  1987, \mn@doi [\mnras] {10.1093/mnras/224.1.13}, 224, 13

\bibitem[\protect\citeauthoryear{Druard et~al.,}{Druard
  et~al.}{2014}]{druard_iram_2014}
Druard C.,  et~al., 2014, \mn@doi [\aap] {10.1051/0004-6361/201423682}, 567,
  A118

\bibitem[\protect\citeauthoryear{Elmegreen \& Scalo}{Elmegreen \&
  Scalo}{2004}]{elmegreen_interstellar_2004}
Elmegreen B.~G.,  Scalo J.,  2004, \mn@doi [\araa]
  {10.1146/annurev.astro.41.011802.094859}, 42, 211

\bibitem[\protect\citeauthoryear{Fan \& {de Grijs}}{Fan \& {de
  Grijs}}{2014}]{fan_star_2014}
Fan Z.,  {de Grijs} R.,  2014, \mn@doi [\apjs] {10.1088/0067-0049/211/2/22},
  211, 22

\bibitem[\protect\citeauthoryear{Federrath, Klessen  \& Schmidt}{Federrath
  et~al.}{2009}]{federrath_fractal_2009}
Federrath C.,  Klessen R.~S.,   Schmidt W.,  2009, \mn@doi [\apj]
  {10.1088/0004-637X/692/1/364}, 692, 364

\bibitem[\protect\citeauthoryear{Gelman \& Rubin}{Gelman \&
  Rubin}{1992}]{gelman_inference_1992}
Gelman A.,  Rubin D.~B.,  1992, \mn@doi [Stat. Science]
  {10.1214/ss/1177011136}, 7, 457

\bibitem[\protect\citeauthoryear{Geyer}{Geyer}{1991}]{geyer_markov_1991}
Geyer C.~J.,  1991. Interface Foundation of North America, \url
  {http://conservancy.umn.edu/handle/11299/58440}

\bibitem[\protect\citeauthoryear{Geyer \& Møller}{Geyer \&
  Møller}{1994}]{geyer_simulation_1994}
Geyer C.~J.,  Møller J.,  1994, Scand. J. Stat., 21, 359

\bibitem[\protect\citeauthoryear{Girichidis, Federrath, Allison, Banerjee  \&
  Klessen}{Girichidis et~al.}{2012}]{girichidis_importance_2012}
Girichidis P.,  Federrath C.,  Allison R.,  Banerjee R.,   Klessen R.~S.,
  2012, \mn@doi [\mnras] {10.1111/j.1365-2966.2011.20250.x}, 420, 3264

\bibitem[\protect\citeauthoryear{Goldstein, Haran, Simeonov, Fricks  \&
  Chiaromonte}{Goldstein et~al.}{2014}]{goldstein_attraction-repulsion_2014}
Goldstein J.~I.,  Haran M.,  Simeonov I.~B.,  Fricks J.,   Chiaromonte F.,
  2014, \mn@doi [Biometrics] {10.1111/biom.12267}, 71, 376

\bibitem[\protect\citeauthoryear{Grasha et~al.,}{Grasha
  et~al.}{2015}]{grasha_spatial_2015}
Grasha K.,  et~al., 2015, \mn@doi [\apj] {10.1088/0004-637X/815/2/93}, 815, 93

\bibitem[\protect\citeauthoryear{Grasha et~al.,}{Grasha
  et~al.}{2017}]{grasha_hierarchical_2017}
Grasha K.,  et~al., 2017, \mn@doi [\apj] {10.3847/1538-4357/aa6f15}, 840, 113

\bibitem[\protect\citeauthoryear{Grasha et~al.,}{Grasha
  et~al.}{2019}]{grasha_spatial_2019}
Grasha K.,  et~al., 2019, \mn@doi [\mnras] {10.1093/mnras/sty3424}, 483, 4707

\bibitem[\protect\citeauthoryear{{Gratier} et~al.,}{{Gratier}
  et~al.}{2017}]{gratier_2017}
{Gratier} P.,  et~al., 2017, \mn@doi [\aap] {10.1051/0004-6361/201629300},
  \href {https://ui.adsabs.harvard.edu/abs/2017A&A...600A..27G} {600, A27}

\bibitem[\protect\citeauthoryear{Guszejnov, Hopkins  \& Krumholz}{Guszejnov
  et~al.}{2017}]{guszejnov_protostellar_2017}
Guszejnov D.,  Hopkins P.~F.,   Krumholz M.~R.,  2017, \mn@doi [\mnras]
  {10.1093/mnras/stx725}, 468, 4093

\bibitem[\protect\citeauthoryear{Haario, Saksman  \& Tamminen}{Haario
  et~al.}{2001}]{haario_adaptive_2001}
Haario H.,  Saksman E.,   Tamminen J.,  2001, Bernoulli, 7, 223

\bibitem[\protect\citeauthoryear{Hollyhead, Bastian, Adamo, Silva-Villa, Dale,
  Ryon  \& Gazak}{Hollyhead et~al.}{2015}]{hollyhead_studying_2015}
Hollyhead K.,  Bastian N.,  Adamo A.,  Silva-Villa E.,  Dale J.,  Ryon J.~E.,
  Gazak Z.,  2015, \mn@doi [\mnras] {10.1093/mnras/stv331}, 449, 1106

\bibitem[\protect\citeauthoryear{Hopkins, Narayanan  \& Murray}{Hopkins
  et~al.}{2013}]{hopkins_meaning_2013}
Hopkins P.~F.,  Narayanan D.,   Murray N.,  2013, \mn@doi [\mnras]
  {10.1093/mnras/stt723}, 432, 2647

\bibitem[\protect\citeauthoryear{Högmander \& Särkkä}{Högmander \&
  Särkkä}{1999}]{hogmander_multitype_1999}
Högmander H.,  Särkkä A.,  1999, Biometrics, 55, 1051

\bibitem[\protect\citeauthoryear{Isham}{Isham}{1984}]{isham_multitype_1984}
Isham V.,  1984, Proceedings of the Royal Society of London. Series A,
  Mathematical and Physical Sciences, 391, 39

\bibitem[\protect\citeauthoryear{Ising}{Ising}{1925}]{ising_beitrag_1925}
Ising E.,  1925, \mn@doi [Zeitschrift für Physik] {10.1007/BF02980577}, 31,
  253

\bibitem[\protect\citeauthoryear{Kennicutt \& Evans}{Kennicutt \&
  Evans}{2011}]{kennicutt_star_2011}
Kennicutt R.~C.,  Evans N.~J.,  2011, \mn@doi [\araa]
  {10.1146/annurev-astro-081811-125610}, 50, 531

\bibitem[\protect\citeauthoryear{Koch \& Rosolowsky}{Koch \&
  Rosolowsky}{2015}]{koch_filament_2015}
Koch E.~W.,  Rosolowsky E.~W.,  2015, \mn@doi [\mnras] {10.1093/mnras/stv1521},
  452, 3435

\bibitem[\protect\citeauthoryear{Kruijssen et~al.,}{Kruijssen
  et~al.}{2019}]{kruijssen_fast_2019}
Kruijssen J. M.~D.,  et~al., 2019, \mn@doi [Nature]
  {10.1038/s41586-019-1194-3}, 569, 519

\bibitem[\protect\citeauthoryear{Krumholz}{Krumholz}{2014}]{krumholz_big_2014}
Krumholz M.~R.,  2014, \mn@doi [\physrep] {10.1016/j.physrep.2014.02.001}, 539,
  49

\bibitem[\protect\citeauthoryear{Kuznetsova, Hartmann  \&
  Ballesteros-Paredes}{Kuznetsova et~al.}{2018}]{kuznetsova_kinematics_2018}
Kuznetsova A.,  Hartmann L.,   Ballesteros-Paredes J.,  2018, \mn@doi [\mnras]
  {10.1093/mnras/stx2480}, 473, 2372

\bibitem[\protect\citeauthoryear{Leininger \& Gelfand}{Leininger \&
  Gelfand}{2017}]{leininger_bayesian_2017}
Leininger T.~J.,  Gelfand A.~E.,  2017, \mn@doi [Bayesian Analysis]
  {10.1214/15-BA985}, 12, 1

\bibitem[\protect\citeauthoryear{Liang}{Liang}{2010}]{liang_double_2010}
Liang F.,  2010, \mn@doi [J. Stat. Comp. Sim.] {10.1080/00949650902882162}, 80,
  1007

\bibitem[\protect\citeauthoryear{Magrini, Stanghellini  \& Villaver}{Magrini
  et~al.}{2009}]{magrini_planetary_2009}
Magrini L.,  Stanghellini L.,   Villaver E.,  2009, \mn@doi [\apj]
  {10.1088/0004-637X/696/1/729}, 696, 729

\bibitem[\protect\citeauthoryear{{Magrini}, {Stanghellini}, {Corbelli}, {Galli}
   \& {Villaver}}{{Magrini} et~al.}{2010}]{magrini_2010}
{Magrini} L.,  {Stanghellini} L.,  {Corbelli} E.,  {Galli} D.,   {Villaver} E.,
   2010, \mn@doi [\aap] {10.1051/0004-6361/200913564}, \href
  {https://ui.adsabs.harvard.edu/abs/2010A&A...512A..63M} {512, A63}

\bibitem[\protect\citeauthoryear{McLaughlin \& Pudritz}{McLaughlin \&
  Pudritz}{1996}]{mclaughlin_formation_1996}
McLaughlin D.~E.,  Pudritz R.~E.,  1996, \mn@doi [\apj] {10.1086/176754}, 457,
  578

\bibitem[\protect\citeauthoryear{Murray, Ghahramani  \& MacKay}{Murray
  et~al.}{2006}]{murray_mcmc_2006}
Murray I.,  Ghahramani Z.,   MacKay D. J.~C.,  2006, in Proceedings of the
  {Twenty}-{Second} {Conference} on {Uncertainty} in {Artificial}
  {Intelligence}. {UAI}'06.
AUAI Press, Arlington, Virginia, United States, pp 359--366

\bibitem[\protect\citeauthoryear{Møller \& Waagepetersen}{Møller \&
  Waagepetersen}{2003}]{moller_statistical_2003}
Møller J.,  Waagepetersen R.,  2003, Statistical {Inference} and {Simulation}
  for {Spatial} {Point} {Processes}.
Chapman \& {Hall}/{CRC} {Monographs} on {Statistics} and {Applied}
  {Probability}, CRC Press

\bibitem[\protect\citeauthoryear{Park \& Haran}{Park \&
  Haran}{2018}]{park_bayesian_2018}
Park J.,  Haran M.,  2018, \mn@doi [J. Am. Stat. Soc.]
  {10.1080/01621459.2018.1448824}, 113, 1372

\bibitem[\protect\citeauthoryear{Peebles}{Peebles}{1980}]{peebles_large-scale_1980}
Peebles P. J.~E.,  1980, The large-scale structure of the universe.
Princeton series in physics, Princeton University Press, Princeton, N.J

\bibitem[\protect\citeauthoryear{Peebles}{Peebles}{1993}]{peebles_principles_1993}
Peebles P. J.~E.,  1993, Principles of {Physical} {Cosmology}.
Princeton University Press

\bibitem[\protect\citeauthoryear{Peebles}{Peebles}{2001}]{peebles_galaxy_2001}
Peebles P. J.~E.,  2001, in {ASP} {Conference} {Proceedings}. San Francisco:
  Astronomical Society of the Pacific, p.~201

\bibitem[\protect\citeauthoryear{Plummer}{Plummer}{1911}]{plummer_problem_1911}
Plummer H.~C.,  1911, \mn@doi [\mnras] {10.1093/mnras/71.5.460}, 71, 460

\bibitem[\protect\citeauthoryear{{R Core Team}}{{R Core Team}}{2018}]{R_2018}
{R Core Team} 2018, R: A Language and Environment for Statistical Computing.
R Foundation for Statistical Computing, Vienna, Austria, \url
  {https://www.R-project.org/}

\bibitem[\protect\citeauthoryear{Ripley \& Kelly}{Ripley \&
  Kelly}{1977}]{ripley_markov_1977}
Ripley B.~D.,  Kelly F.~P.,  1977, \mn@doi [J. London Math. Soc.]
  {10.1112/jlms/s2-15.1.188}, s2-15, 188

\bibitem[\protect\citeauthoryear{Roberts \& Rosenthal}{Roberts \&
  Rosenthal}{2009}]{roberts_examples_2009}
Roberts G.~O.,  Rosenthal J.~S.,  2009, \mn@doi [J. comp. Graph. Stat.]
  {10.1198/jcgs.2009.06134}, 18, 349

\bibitem[\protect\citeauthoryear{Rogers \& Pittard}{Rogers \&
  Pittard}{2013}]{rogers_feedback_2013}
Rogers H.,  Pittard J.~M.,  2013, \mn@doi [\mnras] {10.1093/mnras/stt255}, 431,
  1337

\bibitem[\protect\citeauthoryear{Rosenthal}{Rosenthal}{2011}]{brooks_optimal_2011}
Rosenthal J.,  2011, in Brooks S.,  Gelman A.,  Jones G.,   Meng X.-L.,  eds, ,
  Vol. 20116022, Handbook of {Markov} {Chain} {Monte} {Carlo}.
Chapman and Hall/CRC, \mn@doi{10.1201/b10905-5}

\bibitem[\protect\citeauthoryear{Sharma}{Sharma}{2017}]{sharma_markov_2017}
Sharma S.,  2017, \mn@doi [\araa] {10.1146/annurev-astro-082214-122339}, 55,
  213

\bibitem[\protect\citeauthoryear{Sharma, Corbelli, Giovanardi, Hunt  \&
  Palla}{Sharma et~al.}{2011}]{sharma_population_2011}
Sharma S.,  Corbelli E.,  Giovanardi C.,  Hunt L.~K.,   Palla F.,  2011,
  \mn@doi [\aap] {10.1051/0004-6361/201117812}, 534, A96

\bibitem[\protect\citeauthoryear{Vega, Sánchez  \& Combes}{Vega
  et~al.}{1996}]{vega_self-gravity_1996}
Vega H. J.~d.,  Sánchez N.,   Combes F.,  1996, \mn@doi [Nature]
  {10.1038/383056a0}, 383, 56

\makeatother
\end{thebibliography}



\onecolumn

\appendix

\section{Stability Criteria for GPP Models}
\label{sec:stability}

To construct a well-defined GPP model, crucial stability conditions on GPP are required \citep{moller_statistical_2003}. To provide the definition of the stability criteria, we first need the notion of conditional intensity of a GPP \citep{moller_statistical_2003, baddeley_spatial_2015}, defined as the contribution to the likelihood when adding a point $s$ to the existing pattern $\mathbf{x}$:
\begin{equation}
    \lambda(s, \mathbf{x}) = \frac{f(\mathbf{x} \cup s)}{f(\mathbf{x})}.
\end{equation}
The conditional intensity is an alternative way to define a GPP model to specifying the full probability density function. Given mild condition (hereditary condition \citep{baddeley_spatial_2007}) on the GPP, there is a one-to-one correspondence between the conditional intensity and the probability density function of GPP.

A GPP $\mathbf{X}$ is called Ruelle stable if there exists some positive function $\phi(\cdot)$ defined on $S$ such that $\int_S\phi(s)ds < \infty$ and some constant $c > 0$, satisfying
\begin{equation}
    f(\mathbf{x}) \leq c\prod_{x\in\mathbf{x}}\phi(x)
\end{equation}
for all possible configuration $\mathbf{x}$. $\mathbf{X}$ is locally stable if the conditional intensity satisfies
\begin{equation}
    \lambda(s, \mathbf{x}) \leq \phi(s).
\end{equation}
Local stability implies Ruelle stability. Local stability prevents massive clumping behaviour within a small region when we simulate a GPP, hence, ensuring the existence of simulated point pattern from a given GPP model. Ruelle stability ensures that the GPP model is dominated by a Poisson process, i.e., the probability density function is integrable.

\section{Computation Algorithms} \label{algorithms}

\subsection{Birth-Death Metropolis-Hastings (BDMH) Algorithm for Simulating GPP} \label{simulation}

The algorithm for simulating a GPP is a variant of the Markov chain Monte Carlo (MCMC) algorithms \citep{sharma_markov_2017} called the Birth-Death Metropolis-Hastings (BDMH) algorithm \citep{geyer_simulation_1994}. Given an unnormalised density $h(\mathbf{x})$ of a GPP model, the BDMH attempts to simulate a point pattern from a probability density determined by $h(\mathbf{x})$ through a Markov chain. The state of the Markov chain at each time step is a point pattern; we denote it by $\mathbf{X}_1, \mathbf{X}_2, ..., \mathbf{X}_t$. At each $t$, a point is either added (``born") to the point pattern with probability $p_b$ or removed (``dies") from the point pattern with probability $p_d = 1- p_b$. If a point is to be born, it is selected according to some arbitrary probability density $b(\mathbf{X}_t; s)$ over the observation window where $s$ is the newly added point; If a point is to be removed, it is selected with another arbitrary probability density $d(\mathbf{X}_t; s)$ on the existing point pattern where $s$ is the point to be removed. Lastly, we calculate the acceptance probability for the proposal and determine whether it is accepted or not.

Here we illustrate the construction of BDMH algorithm for simulating a point pattern from a GPP model. In general, let $\mathbf{X}$ be a GPP with unnormalised probability density $h(\cdot)$. To formalise the algorithm, let 
\begin{equation}
    \mathbf{X}^+ = \mathbf{X}_t\cup \{s\}
\end{equation}
be the point pattern formed when adding $s$ into $\mathbf{X}_t$ and
\begin{equation}
    \mathbf{X}^- = \mathbf{X}_t\setminus \{s\}
\end{equation}
be the point pattern formed when removing $s$ from $\mathbf{X}_t$. The algorithm proceeds as follows:

\vspace{2mm}
\begin{algorithm*}[H]
\label{BDMH}
\SetAlgoLined
Input: Initial point pattern $\mathbf{X}_0$, number of iterations $T$, birth-move probability $p_b$, birth density $b(\cdot; \cdot)$, death density $d(\cdot;\cdot)$\;
 \For{t = 1,...,$T$}{
  Draw $U \sim \text{unif}(0,1)$\;
  \eIf{$U < p_b$,}{
   Generate $s \sim b(\mathbf{X}_t;s)$\;
   Calculate $
   r_b = \dfrac{h(\mathbf{X}^+)d(\mathbf{X}^+;s)p_d}{h(\mathbf{X}_t)b(\mathbf{X}_t;s)p_b}
   $\;
   Accept $\mathbf{X}^+$ with probability $a_b = \min(1, r_b)$;
   }{
   Select $s \sim d(\mathbf{X}_t;s)$ from $\mathbf{X}_t$\;
   Calculate $
   r_d = \dfrac{h(\mathbf{X}^-)b(\mathbf{X}^-;s)p_b}{h(\mathbf{X}_t)d(\mathbf{X}_t;s)p_d}
   $\;
   Accept $\mathbf{X}^-$ with probability $a_d = \min(1, r_d)$;
   }
  }
 \caption{Birth-Death Metropolis-Hastings Algorithm}
\end{algorithm*}

The hyper-parameters of the algorithms, such as $b(\mathbf{X}_t; s)$, can have drastic effects on the convergence of the algorithm. We here specify our scheme for choosing hyperparameters. For simplicity, we choose the birth move probability $p_b = p_d = 0.5$. For the birth probability density $b(\mathbf{X}_t; s)$, we set it as the following:
\begin{equation}
\label{birth_density}
    b(\mathbf{X}_t; s) \propto \left(1 + \frac{d^2(s, \mathbf{x}_G)}{h^2}\right)^{-1},
\end{equation}
where $ d(s, \mathbf{x}_G)$ is the distance from a point $s \in S$ to the closest GMCs. $h$ is a hyperparameter. The motivation behind this choice of birth density is that majority of YSCCs in the data are very close to GMCs. Therefore, we need to adjust the birth probability density so that generating a point close to a GMC is reasonably probable. Otherwise, the chain can propose points that are too far from any GMCs,  these proposals will all get rejected, and the simulation will take a long time to converge to resemble the data. However, there is no built in method to generate the points from the probability density given by equation \ref{birth_density}, hence, we resort to a simple rejection method and generate 200,000 points that follow the distribution. We then select a point from this sample uniformly for each birth proposal, and the generated point will then follow the distribution specified by \ref{birth_density}. The value of $h$ is set to $0.01$ based on a visual inspection of the generated sample from the birth distribution where the points distribute in a similar fashion to that in the data. Furthermore, there exists a normalizing constant for the birth probability density which is required for calculating the density; we obtain it by splitting the observation window into a 500$\times$500 grid and calculate the unnomarlized density value for each grid point. We then sum up the values for all grid points as an approximation for the normalizing constant. For the death probability density, we let $d(\mathbf{X}_t; s) = 1/n(\mathbf{X}_t)$, i.e., uniformly choosing a point for removal. 

\subsection{Bayesian Inference for GPP models} \label{inference}

Inference for GPPs is generally a daunting task. Inference through maximum likelihood estimation (MLE) and Bayesian approaches both exist, but MLE approaches such as the maximum pseudo-likelihood estimation (MPLE) require the model to be in log-linear form \citep{baddeley_practical_2000} while Markov chain Monte Carlo MLE \citep{geyer_markov_1991} requires the model to have analytical gradients and be numerically stable. These requirements are very restrictive when formulating the model. Furthermore, we believe the Bayesian approach is much more suitable for our purpose as future observations and acquisition of new data embodies the concept of information update which is naturally incorporated in the Bayesian paradigm.

However, the standard method for Bayesian inference such as Metropolis-Hastings (MH) algorithm is not feasible for our model. This is because our likelihood function itself contains an unnormalised constant $\alpha$ as mentioned before, which is a function of the parameters. Several methods have been proposed to deal with this issue. In this paper, we will adopt the method proposed by \cite{liang_double_2010} called the double Metropolis-Hastings (DMH) algorithm which originated from the exchange algorithm by \cite{murray_mcmc_2006}. 

Here we discuss some of the ideas and constructions of Markov chain Monte Carlo algorithms for GPP models. In general, a GPP model can be written as 
\begin{equation}
    f(\mathbf{x}; \boldsymbol{\theta}) = \frac{h(\mathbf{x}| \boldsymbol{\theta})}{\alpha(\boldsymbol{\theta})},
\end{equation}
where $h(\mathbf{x}| \boldsymbol{\theta})$ is the part (unnormalised) that we can define and $\alpha(\boldsymbol{\theta})$ is an intractable normalising constant which is a function of the parameters $\boldsymbol{\theta}$. $f$ in this case is called a doubly-intractable distribution \citep{murray_mcmc_2006, park_bayesian_2018}.

Assuming prior distribution $p(\boldsymbol{\theta})$ and the proposal distribution $q(\boldsymbol{\theta}'|\boldsymbol{\theta})$, the posterior distribution is then
\begin{equation}
    \pi(\boldsymbol{\theta}|\mathbf{x}) \propto f(\mathbf{x}|\boldsymbol{\theta})p(\boldsymbol{\theta}).
\end{equation}
Carrying out the standard MCMC algorithm is simple when $f(\cdot)$ is known. However, for a doubly-intractable distribution, the problem arises when we calculate the Metropolis-Hastings ratio
\begin{equation}
    r = \frac{f(\mathbf{x}|\boldsymbol{\theta}')p(\boldsymbol{\theta}')q(\boldsymbol{\theta}|\boldsymbol{\theta}')}{f(\mathbf{x}|\boldsymbol{\theta})p(\boldsymbol{\theta})q(\boldsymbol{\theta}'|\boldsymbol{\theta})} = \frac{h(\mathbf{x}| \boldsymbol{\theta}')p(\boldsymbol{\theta}')q(\boldsymbol{\theta}|\boldsymbol{\theta}')\alpha(\boldsymbol{\theta})}{h(\mathbf{x}| \boldsymbol{\theta})p(\boldsymbol{\theta})q(\boldsymbol{\theta}'|\boldsymbol{\theta})\alpha(\boldsymbol{\theta}')}
\end{equation}
where the ratio
\begin{equation*}
\centering
    \frac{\alpha(\boldsymbol{\theta})}{\alpha(\boldsymbol{\theta}')}
\end{equation*}
is unknown. This makes the acceptance ratio in a MH-update unavailable to us and normal MCMC sampling cannot proceed. We present some of the existing methods that deal with this issue. 
The main idea of the algorithms is to simulate an auxiliary variable to remove the unknown normalising constant ratio and render the inference feasible.

The first such algorithm is called the exchange algorithm, proposed by \cite{murray_mcmc_2006} and illustrated in Algorithm \ref{exchange}.
\vspace{2mm}
\begin{algorithm}[H] 
\label{exchange}
\SetAlgoLined
Input: Initial $\boldsymbol{\theta}$, number of iterations $T$\;
\For{
$t = 1,...,T$
}{
Propose $\boldsymbol{\theta}' \sim q(\boldsymbol{\theta}'|\boldsymbol{\theta})$\;
Generate auxiliary variable $\mathbf{y}\sim h(\cdot|\boldsymbol{\theta}')/\mathcal{Z}(\boldsymbol{\theta}')$\;
Calculate $r = \dfrac{h(\mathbf{x}| \boldsymbol{\theta}')h(\mathbf{y}| \boldsymbol{\theta})p(\boldsymbol{\theta}')q(\boldsymbol{\theta}|\boldsymbol{\theta}')}{h(\mathbf{x}| \boldsymbol{\theta})h(\mathbf{y}| \boldsymbol{\theta}')p(\boldsymbol{\theta})q(\boldsymbol{\theta}'|\boldsymbol{\theta})}$\;
Accept $\boldsymbol{\theta}'$ with probability $a = \min(1, r)$;
}
\caption{Exchange Algorithm}
\end{algorithm}
\vspace{2mm}

The exchange algorithm essentially introduces an auxiliary variable $\mathbf{y} \sim h(\cdot|\boldsymbol{\theta}')/\alpha(\boldsymbol{\theta}')$ so that the ratio between the normalising constants vanishes. Another way to understand the algorithm is that the unknown ratio $\alpha(\boldsymbol{\theta})/\alpha(\boldsymbol{\theta}')$ is approximated by its unbiased estimator $h(\mathbf{y}| \boldsymbol{\theta})/h(\mathbf{y}| \boldsymbol{\theta}')$. Surprisingly, this substitute of estimator leads to an asymptotically exact algorithm \citep{murray_mcmc_2006, park_bayesian_2018}. The drawback, however, is that it requires $\mathbf{y}$ following $h(\cdot|\boldsymbol{\theta}')/\alpha(\boldsymbol{\theta}')$ perfectly to ensure the algorithm is asymptotically exact. The perfect simulation requirement is prohibitive for most GPP models since its construction for more complex models is unknown or impossible \citep{park_bayesian_2018}. \cite{liang_double_2010} proposed a double Metropolis-Hastings (DMH) algorithm to relax the perfect sampling restriction so that the computation becomes feasible. The DMH algorithm is illustrated in Algorithm \ref{DMH}.

\vspace{2mm}
\begin{algorithm}[H]
\label{DMH}
\SetAlgoLined
Input: Initial $\boldsymbol{\theta}$, number of iterations $T$, number of iterations $M$ for simulating auxiliary variable through BDMH\;
\For{
$t = 1,...,T$
}{
Propose $\boldsymbol{\theta}' \sim q(\boldsymbol{\theta}'|\boldsymbol{\theta})$\;
Generate auxiliary variable $\mathbf{y}\sim h(\cdot|\boldsymbol{\theta}')/\mathcal{Z}(\boldsymbol{\theta}')$\ through $M$-step BDMH algorithm\;
Calculate $r = \dfrac{h(\mathbf{x}| \boldsymbol{\theta}')h(\mathbf{y}| \boldsymbol{\theta})p(\boldsymbol{\theta}')q(\boldsymbol{\theta}|\boldsymbol{\theta}')}{h(\mathbf{x}| \boldsymbol{\theta})h(\mathbf{y}| \boldsymbol{\theta}')p(\boldsymbol{\theta})q(\boldsymbol{\theta}'|\boldsymbol{\theta})}$\;
Accept $\boldsymbol{\theta}'$ with probability $a = \min(1, r)$;
}
\caption{Double Metropolis-Hastings (DMH) Algorithm}
\end{algorithm}
\vspace{2mm}

The idea is to simply replace the perfect simulation of the auxiliary variable $\mathbf{y}$ by a BDMH simulation of the point pattern from $h(\cdot|\boldsymbol{\theta}')/\alpha(\boldsymbol{\theta}')$. This relaxation leads to an algorithm that is asymptotically non-exact since BDMH can only provide an approximate simulation of point patterns following the target distribution. This problem can be circumvented by running the BDMH simulation sufficiently long, usually 50~100 times the number of points in the point pattern, to reduce the approximation error. Although this might increase the computational burden, modern supercomputer and multi-core processing are powerful enough to render the computation feasible. In this paper, the DMH algorithm will be used to carry out the inference.

\section{Convergence Diagnostics of DMH Algorithm} \label{convergence_daignose}

\begin{figure*}
\centering
\includegraphics[width = 175mm]{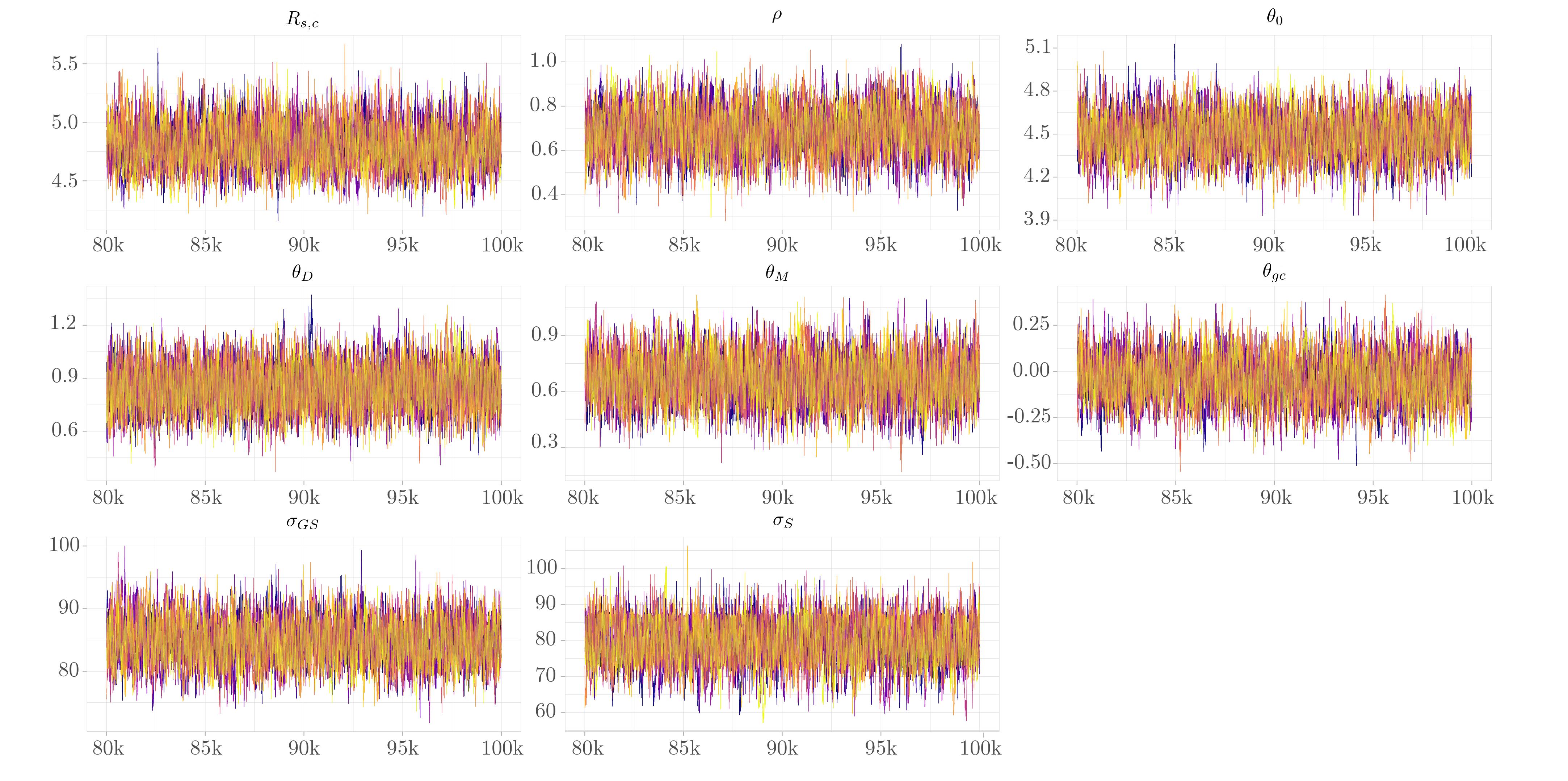}
\caption[Traceplot of each model parameter obtained from ten MCMC runs.]{Traceplot of each model parameter obtained from ten MCMC runs for 100k iterations. The plot only shows the last 20k iterations for improved visualization.}
\label{traceplot}
\end{figure*}

Figure \ref{traceplot} shows the traceplots of 10 independently run MCMC chains of 50k iterations obtained from the DMH algorithm. The plots only show the last 20k iteration for better visualisation. The Gelman-Rubin convergence statistic \citep{gelman_inference_1992} is well below 1.001, indicating the convergence of the chains.  

\bsp	
\label{lastpage}
\end{document}